\RequirePackage{rotating}
\documentclass[fleqn,usenatbib]{mnras}

\usepackage{newtxtext,newtxmath}

\usepackage[T1]{fontenc}
\usepackage{ae,aecompl}

\usepackage{graphicx,rotating}	
\usepackage{amsmath}	
\usepackage{amssymb}	
\usepackage[dvipsnames]{xcolor}

\graphicspath{ {figures/} }
\newcommand{\hda}{H$\delta_{\rm A}$}
\newcommand{\dn}{Dn4000}

\definecolor{mycolor}{rgb}{0,0,0}

\title[Post-starburst galaxy star formation histories]{The star formation histories of z$\sim1$ post-starburst galaxies}

\author[Wild et~al.]
{Vivienne Wild$^{1}$\thanks{E-mail: \textcolor{mycolor}{vw8@st-andrews.ac.uk}},
Laith Taj Aldeen$^{1,2}$,
Adam Carnall$^3$,
David Maltby$^4$,
\newauthor
Omar Almaini$^4$,
Ariel Werle$^{5,6}$,
Aaron Wilkinson$^{1,7},$
Kate Rowlands$^8$,
\newauthor
Micol Bolzonella$^9$,
Marco Castellano$^{10}$,
Adriana Gargiulo$^{11}$,
Ross McLure$^{3}$,
\newauthor
Laura Pentericci$^{10}$,
Lucia Pozzetti$^{9}$
\\$^{1}$ SUPA\thanks{Scottish Universities Physics Alliance}, School of Physics \& Astronomy, University of St Andrews, North Haugh, St Andrews, Fife KY16 9SS, UK
\\$^{2}$Department of Physics, College of Science, University of Babylon, Hillah, Babylon, P.O. Box 4, Iraq. 
\\$^{3}$ SUPA\, Institute for Astronomy, University of Edinburgh, Royal Observatory, Edinburgh EH9 3HJ, UK
\\$^{4}$University of Nottingham, School of Physics and Astronomy, Nottingham NG7 2RD, UK
\\$^{5}$Instituto de Astronomia, Geof\'{\i}sica e Ci\^{e}ncias Atmosf\'{e}ricas, Universidade de S\~{a}o Paulo, R. do Mat\~{a}o 1226, 05508-090 S\~{a}o Paulo\, Brazil
\\$^{6}$Departamento de F\'isica - CFM - Universidade Federal de Santa Catarina, Florian\'opolis, SC, Brazil
\\$^7$ Universiteit Gent, Sterrenkundig Observatorium, Gent, Belgium
\\$^{8}$ Space Telescope Science Institute, 3700 San Martin Drive, Baltimore, MD 21218, USA
\\$^{9}$ INAF - Osservatorio di Astrofisica e Scienza dello Spazio di Bologna via Gobetti 93/3, 40129 Bologna, Italy
\\$^{10}$ INAF - Osservatorio Astronomico di Roma, Via Frascati 33, 00078 Monte Porzio Catone, RM, Italy
\\$^{11}$ INAF - IASF, Via Alfonso Corti 12, I-20133 Milano, Italy
}

\date{Accepted XXX. Received YYY; in original form ZZZ}

\pubyear{2020}

\begin{document}
\label{firstpage}
\pagerange{\pageref{firstpage}--\pageref{lastpage}}
\maketitle

\begin{abstract}
We present the star formation histories of 39 galaxies with high quality rest-frame optical spectra at $0.5<z<1.3$ selected to have strong Balmer absorption lines and/or Balmer break, and compare to a sample of spectroscopically selected quiescent galaxies at the same redshift. Photometric selection identifies a majority of objects that have clear evidence for a recent short-lived burst of star formation within the last 1.5\,Gyr, i.e. ``post-starburst'' galaxies, however we show that good quality continuum spectra are required to obtain physical parameters such as burst mass fraction and burst age. Dust attenuation appears to be the primary cause for misidentification of post-starburst galaxies, leading to contamination in spectroscopic samples where only the [OII] emission line is available, as well as a small fraction of objects lost from photometric samples. The 31 confirmed post-starburst galaxies have formed 40-90\% of their stellar mass in the last 1-1.5\,Gyr. We use the derived star formation histories to find that the post-starburst galaxies are visible photometrically for 0.5-1\,Gyr. This allows us to update a previous analysis to suggest that 25-50\% of the growth of the red sequence at z$\sim$1 could be caused by a starburst followed by rapid quenching. We use the inferred maximum historical star formation rates of several 100-1000\,M$_\odot$/yr and updated visibility times to confirm that sub-mm galaxies are likely progenitors of post-starburst galaxies. The short quenching timescales of 100-200\,Myr are consistent with cosmological hydrodynamic models in which rapid quenching is caused by the mechanical expulsion of gas due to an AGN.

\end{abstract}




\section{Introduction}\label{sec:intro}
Uncovering the physical processes that lead to the increasing number of quiescent galaxies over cosmic time remains a challenge for both observational and theoretical extragalactic astronomy. Aggregated statistics such as stellar mass functions, colour-mass relations and quenched fractions do not provide unique constraints on different possible quenching models \citep[e.g.][]{Skelton2012}. However, by identifying recently quenched galaxies at all redshifts, we can attempt to study their properties in sufficient detail to understand on a case-by-case basis what might have caused the quenching to happen. 

Analysis of both the morphologies and spectral energy distributions (SEDs) of high redshift quiescent galaxies has led to growing evidence for at least two quenching mechanisms that act on ``fast'' and ``slow'' timescales \citep[e.g][]{Pacifici2016,Moutard:2016p10257,Maltby2018,Wu2018,Rowlands2018,Belli2019}.  In the case where the shut-off in star formation is both fast and sufficiently substantial, for example following a short lived burst of star formation, stellar evolution leaves a useful identifiable imprint on the integrated continuum spectra of galaxies for up to $\sim$1\,Gyr, with dominant A and F stars leading to a strong Balmer break and Balmer absorption lines, while the lack of O and B stars leaves little ultraviolet flux or nebular emission. Often termed ``post-starburst'',  ``E+A'' or ``K+A'' galaxies are identified either through their unusual triangular multi-wavelength spectral energy distribution (SED), or through their strong Balmer absorption lines \citep[e.g.][]{Dressler1983}. Throughout this paper we refer to these galaxies as ``post-starburst'', as they are not morphologically selected (as in E+A's) and this more closely relates to the physical processes going on. We note that for sensitive measurements, a rapid shut-off in star formation of a highly star forming galaxy may also cause strong Balmer lines and break, and in these cases ``rapidly quenched'' galaxy may be more appropriate nomenclature \citep{Couch:1987p2487,vonderLinden2010,Pawlik2019}. In this paper, we focus on the observationally easily identified ``rapidly quenched'' galaxies. Are they truly ``post-starburst'' or just ``rapidly quenched''? Can we use their recent star formation history (SFH) to better constrain their contribution to the overall growth of the quiescent population via a rapid quenching mode? And can we pinpoint their progenitors? 

A useful property of post-starburst galaxies is the possibility to age-date the burst by making use of the rapid change of the Balmer break and Balmer absorption line strength with burst age. Balmer absorption line strength alone still leaves us with a degeneracy between the age and strength of the starburst \citep{Wild2007}, however there are additional features in the optical spectra that can break this degeneracy such as the CaII(H\&K) to Balmer line ratio \citep{Leonardi1996}. Spectral fitting of low-redshift post-starburst galaxies has led to useful constraints on the burst strengths and timescales involved in the burst, which help to pin down the cause of the starburst as well as the quenching processes. At low-redshift, burst mass fractions of as much as 70\% strongly implicate major mergers as the cause of the burst in low-redshift post-starburst galaxies, consistent with their morphological features, while the rapid decline rate in star formation, alongside the decline in molecular gas, implicates AGN feedback as a quenching process, at least at high mass \citep{Kaviraj2007,Pawlik2018,French2018}.  

Local post-starbursts have been known for decades to be predominantly elliptical in morphology, with a large fraction showing signs of morphological disturbance, and have long been linked to a transition population between major gas rich mergers and quiescent elliptical galaxies \citep[see][for a review]{Pawlik2018}. The highly compact nature of massive high redshift ($z\gtrsim$1) post-starburst galaxies, and clear dissimilarity from the morphologies of star-forming progenitors, suggests that morphological transformation via dissipative collapse precedes the quenching of star-formation for these objects \citep[e.g.][]{Yano2016,Almaini2017,Wu2018}. However, the prevalence of such extreme events is likely dependant on stellar mass, environment and redshift. Intermediate redshift ($0.5<z<1$) post-starburst galaxies are typically lower mass and less concentrated than their high redshift counterparts \citep{Maltby2018}, and their notable excess in galaxy clusters indicates that environmental processes may lead to the majority of rapidly quenched galaxies at $z<1$ \citep[e.g.][]{Vergani:2010p9474,Socolovsky2018,Moutard:2018,Paccagnella2019,Owers2019}.

The consensus for low-redshift post-starburst galaxies is that a large fraction are transition galaxies, forming the evolutionary link between gas-rich major mergers, ultraluminous infrared galaxies (ULIRGs), and future quiescent elliptical galaxies. However, with high quality data it is possible to identify post-starburst features arising from three separate processes \citep{Pawlik2018}: traditional blue$\rightarrow$red quenching, cyclical evolution within the blue sequence, as well as rejuvenation of red-sequence galaxies. These results are in qualitative agreement with cosmological hydrodynamical simulations \citep{Pawlik2019}, however more work is required to determine whether the relative fraction of the different processes is correctly reproduced in the simulations, especially because different selection methods lead to observed samples with different physical properties \citep{Pawlik2018,French2018}.

At higher redshifts, substantial and rapid shut-offs in star formation may account for a significant fraction of red-sequence growth \citep{Wild2009,Wild2016,Rowlands2018,Forrest2018,Belli2019}, however, in order to determine this fraction more accurately we require constraints on the time for which the post-starburst features are visible (visibility time). \citet{Belli2019} used spectral fitting of rest-frame optical spectra and broad band photometry to estimate the time spent in the post-starburst phase from the median stellar ages of a small number of post-starburst galaxies at $1.5<z<2.5$. In this paper, we go one step further to calculate the visibility times for individual post-starburst galaxies directly from their fitted star formation histories, making use of high quality rest-frame optical spectra. From this we can directly calculate the fraction of red-sequence growth accounted for by this fast post-burst quenching phenomenon, as well as identify likely progenitors.

The outline of the paper is as follows. In Section \ref{sec:data} we introduce our dataset, describe the analysis using the spectral fitting package {\sc bagpipes}, and our sample selection of post-starburst candidates using both photometry and spectroscopy. In Section \ref{sec:results} we show the derived star formation histories of the post-starburst candidates, quantify their burst masses and ages, and investigate the role of dust in causing post-starburst galaxies to be missed from photometric selection methods, and contaminants to arise in spectroscopic samples. In Section \ref{sec:discussion} we discuss our results with respect to the growth of the red sequence and the likely progenitors of the post-starburst galaxies. Finally, we investigate the benefit of high quality continuum spectroscopy over multiwavelength photometric data for estimating the physical properties of post-starburst galaxies. Where necessary we assume a cosmology with  $\Omega_M= 0.3$, $\Omega_\Lambda=0.7$ and $h = 0.7$. All magnitudes are on the AB system \citep{Oke:1983p10013}. Stellar masses are calculated assuming a \citep{Chabrier2003} initial mass function (IMF) and are defined as the stellar mass remaining at the time of observation.

\section{Data and analysis}\label{sec:data}
The UKIRT Infrared Deep Sky Survey (UKIDSS) Ultra Deep Survey
(UDS) Data Release 11 (DR11), is a deep, large area near-infrared (NIR) imaging
survey and the deepest of the UKIDSS surveys
\citep[][Almaini et al. in prep]{Lawrence:2007p9799}. The survey area is 0.63 square degrees, after masking bright stars and cross talk and combining with optical and mid-IR imaging. UKIRT observations provide $J$, $H$, $K$
observations to $5\sigma$ limiting depths in 2\arcsec\ diameter
apertures of 25.4, 24.8 and 25.3 AB magnitudes respectively (Almaini et al. in prep). Deep optical observations come from the
Subaru XMM-Newton Deep Survey \citep[SXDS, ][]{Furusawa:2008p8792}, to
depths of 27.2, 27.0, 27.0 and 26.0 in $V$, $R$, $i'$, $z'$ ($5\sigma$, 2\arcsec). $Y$-band coverage with a depth of 23.9 comes from the VISTA-Video survey (P.I. M.~Jarvis). Mid-IR
coverage (IRAC 3.6$\mu m$ and 4.2$\mu m$) is provided by the Spitzer UDS Legacy Program
(SpUDS, PI:Dunlop) to a depth of 24.2. Photometry was extracted within
2\arcsec\ diameter apertures at the position of the $K$-band sources,
with an aperture correction applied for the IRAC 3.6$\mu m$\ and 4.5$\mu m$ 
images. Further details on the methods used to construct the UDS DR11
catalogue can be found in Almaini et al. in prep. 

Following the method presented in \citet{Wild:2014p9476}, super-colours are calculated for all galaxies with $K<24.5$ and $0.5<z_{\rm phot}<3.0$, using photometric redshifts as described in Almaini et al. (in prep) and Hartley et al. (in prep) or spectroscopic redshifts where available. These super-colours represent a linear combination of observed frame filters, in the same way as traditional colour-colour diagrams. However, the linear combination is optimised to maximise the variance in the dataset using a Principal Component Analysis of model SEDs, and data is not forced to fit model SEDs unlike during a k-correction process. The first and second principal component amplitude (termed SC1 and SC2) are weighted linear combinations of observed-frame fluxes that describe the overall red/blue colour of the SED and the strength of the Balmer or 4000\AA\ break respectively. In DR11, the addition of the $Y$-band data, extension to bluer rest-frame wavelengths of 2500\AA, as well as improvements in the reduction of the IRAC 4.5$\mu m$, have allowed us to extend the redshift range slightly to $0.5<z<3$ (Wilkinson et al. in prep.). For this paper, the only impact is a  slight alteration of the eigenbasis, and therefore the exact values of the super-colours differ a little from those in \citet{Wild2016} as a consequence. The boundaries between the SED classes have been carefully shifted to align with those derived in \citet{Wild:2014p9476}.  

A large number of objects in the UDS field have been observed spectroscopically; from all sources available to us we select galaxies with spectra that have led to secure spectroscopic redshifts and are also included in the DR11 super-colour catalogue. Further details of the spectroscopic datasets are provided in \citet{Maltby2019}, and we give a brief summary here. The UDSz project used a combination of the VIMOS and FORS2 instruments on the ESO VLT to observe galaxies with $K_{AB}<23.0$ (ESO Large Programme 180.A-0776, PI:Almaini) providing 1156 spectra after matching to the DR11 super-colour catalogue. The VANDELS project provides a further 239 from data release 2, observed with the upgraded VIMOS instrument \citep{VANDELS1, VANDELS2}. Finally, 44 additional spectra were observed as part of ESO program 094.A-0410, again with the upgraded VIMOS instrument. These observations primarily targetted super-colour selected post-starbursts (see previous paragraph) and are described in \citet[][DM16 hereafter]{Maltby2016}. The different spectroscopic samples differ slightly in their central wavelengths, spectral resolutions and sampling. Spectral resolutions (FWHM) are 200, 660 and 580 for the UDSz-VIMOS, UDSz-FORS and VANDELS/DM16. Prior to performing our analysis, we binned all three VIMOS samples by 2 pixels, to obtain approximately Nyquist sampling. All spectra were visually inspected and compared to the photometry to identify regions where there were catastrophic problems that would lead to problems during fitting. The first and last 200 pixels were removed from all spectra and those that extended beyond 9500\AA\ were additionally masked beyond that point where the data reduction becomes less reliable. The visual inspection revealed an unphysical drop in flux in the red end of a few of the VANDELS and DM16 spectra, and these few spectra were therefore masked above 9250\AA.  The VANDELS and DM16 spectra were additionally masked below 5000\AA\ due to typically very low signal-to-noise (SNR) ratio. 

\subsection{Spectral fitting}

To aid our spectroscopic sample selection, we perform an initial fit to the combined stellar continuum spectrum and photometric data of all spectroscopically observed galaxies with the {\sc bagpipes} code described in detail in \citet{Carnall2018}, using the same SFH model and priors as in \citet{CarnallVANDELS2019}. {\sc bagpipes} is a fully Bayesian spectral fitting code, that fits observed spectroscopic and photometric SEDs to spectral synthesis models to obtain the probability distribution functions (PDF) for parameters describing the star formation history, dust and metallicity content of each galaxy. For this initial fit, the aim is simply to get a good, physically plausible fit, to both the photometry and spectra.

We use the \citet{BC03} spectral synthesis models, updated to 2016\footnote{\url{http://www.bruzual.org/~gbruzual/bc03/Updated_version_2016/}}, built from both observed and theoretical stellar spectra, assuming a \citet{Chabrier2003} initial mass function and `Padova (1994)' \citep{Alongi.etal.1993a,Bressan.etal.1993a,Fagotto.etal.1994a,Fagotto.etal.1994b,Girardi.etal.1996a} evolutionary tracks. This version of the stellar synthesis models includes the observed MILES stellar library \citep{SanchezBlazquez.etal.2006a,FalconBarroso.etal.2011a} in the wavelength range 3540\AA - 7350\AA, extended with the STELIB stellar library \citep{LeBorgne.etal.2003a} out to 8750\AA. Theoretical spectra complement the observed spectra from the Tlusty \citep{Lanz.Hubeny.2003a, Lanz.Hubeny.2003b}, Martins \citep{Martins.etal.2005a}, UVBlue \citep{UVblue2005}, PoWR \citep{Sander.Hamann.Todt.2012a}, BaSeL 3.1 \citep{Aringer2009}, IRTF \citep{IRTF2009} libraries, as well as dusty TP-AGB stars \citep{Nenkova2000, GonzalezLopezlira2010}, extending the models into the near-infrared and ultraviolet wavelength ranges and covering more unusual spectral types. Even with this extensive combination of theoretical and observed stellar spectra, there are inevitable gaps. Of particular importance for this work on post-starburst galaxies is a lack of A stars included in the models bluewards of 3000\AA. For this reason, as well as the problems caused by an uncertain and potentially varying dust attenuation law and 2175\AA\ dust feature, we mask the spectra bluewards of $<3000$\AA\ and increase the errors on the photometry where the central rest-frame wavelength is bluewards of $<3000$\AA\ to a maximum SNR of 10.

For the initial fit, we assume a standard double power law star formation rate (SFR) as a function of time: 
\begin{equation}
SFR(t) \propto \left[ \left(\frac{t}{\tau}\right)^\alpha + \left(\frac{t}{\tau}\right)^{-\beta} \right]^{-1}  
\end{equation}
where $t$ is time from the formation epoch, $\alpha$ is the rising slope, $\beta$ is the falling slope and $\tau$ determines the position of the peak SFR.  We find that the final sample selection is not sensitive to the precise details of the assumed SFH or other details of the fit. Metallicity is free to vary between 1/100th and 2.5 times solar, but does not evolve with time. A nebular component is included with ionisation parameter $\log U=-3$, and a two component \citet{CF00} dust attenuation with a variable slope for the attenuation curve and attenuation strength, and stars younger than $10^7$ years twice as attenuated as those older. A second order multiplicative polynomial makes allowances for any inaccuracies in the spectrophotometric calibration, while a Gaussian process noise component accounts for systematic correlated noise in the spectra. Both these components are crucial to obtaining a good fit to these $z\sim1$ spectra.  For full details of this initial fit, including ranges and priors on parameters, see \citet{CarnallVANDELS2019}. We additionally carry out a run with a pure white-noise scaling rather than Gaussian process noise, to ensure that this additional feature does not significantly impact the final sample selection.  A Gaussian prior is set on the spectroscopic catalogue redshift, with a $\sigma$-width of 0.01. 

Before performing the fit, we mask regions that might contain emission lines in the rest-frame with a mask of $\pm5$\AA\ (most notably [OII] and H$\delta$ in the observed wavelength range) and observed-frame atmospheric telluric features between 7580--7650\AA. Although {\sc bagpipes} does simultaneously fit emission lines this prevents confusion where emission lines are caused by non-stellar processes as may be common in post-starburst galaxies \citep{Yan2006,Wild2007,Wild2010,Alatalo2016}. We verified that our results are not significantly altered when lines are not masked.  The errors on the photometry are set to a maximum SNR of 20, or 10 for the IRAC bands or where the central rest-frame wavelength of the filter is $<3000$\AA\ (see above). The UDSz and VIMOS spectra were corrected for Galactic extinction using interpolated \citet{SFD98} maps and the \citet{CCM88} extinction law. The reduced VANDELS spectra are already corrected for Galactic extinction \citep{VANDELS2}.

While the quality of the spectra are generally extremely high for these redshifts, we found that the limited signal-to-noise as well as various spectrophotometric inaccuracies limited our ability to select a robust sample of post-starburst galaxies from spectral lines measured in the raw spectra. We therefore measured the \hda\ \citep{HdA} and \dn\ \citep{Dn4000} spectral indices from the fitted stellar continuum. This has the added advantage of excluding the emission lines from our spectral indices, making our measurements more representative of the recent star formation history. We note that the \hda\ index values are lower by $\sim$1\AA\ than the H$\delta$ equivalent width used in some other post-starburst work \citep[e.g.][]{Goto2007,Maltby2016}, after allowing for the fact that these works often include the infilled emission lines in their measurements. We choose to use \hda\ for consistency with more general work on galaxy populations with the Sloan Digital Sky Survey \citep[e.g.][]{Kauffmann2003} and more recent analyses at higher redshift \citep[e.g.][]{Wu2018}. We measure the [OII] line equivalent width (W[OII]) directly from the spectra, but use the fitted model continuum to define the continuum above which to integrate, and integrate between $\pm3.5\times \sigma_{disp}$ where $\sigma_{disp}$ is the width of the Gaussian kernel used to convolve the spectrum during the fit\footnote{We note this is not the velocity dispersion of the galaxy, as it does not account for the resolution of the spectrum or the models.}. 

When analysing both spectra and photometry together we must be aware of possible aperture effects due to stellar population radial gradients in the galaxies. The photometry is extracted in 2\arcsec\ apertures, while the slitwidths are typically 1\arcsec\ and may therefore miss a fraction of the outer regions of the galaxies. For the post-starburst galaxies, their relatively compact effective radii of 0.2-0.3\arcsec\ and negligible radial colour gradients \citep{Maltby2018} indicate aperture effects will be minimal. However, in the case of younger outer regions it is possible that some nuclear post-starbursts present in the spectroscopic sample will not be well fit by our combined spectra and photometry fitting method, due to excess light from outer young stars dominating the shape of the photometric SED. 

\subsubsection{Two component model for post-starburst galaxies}

\begin{figure}
    \centering
    \includegraphics[width=\columnwidth]{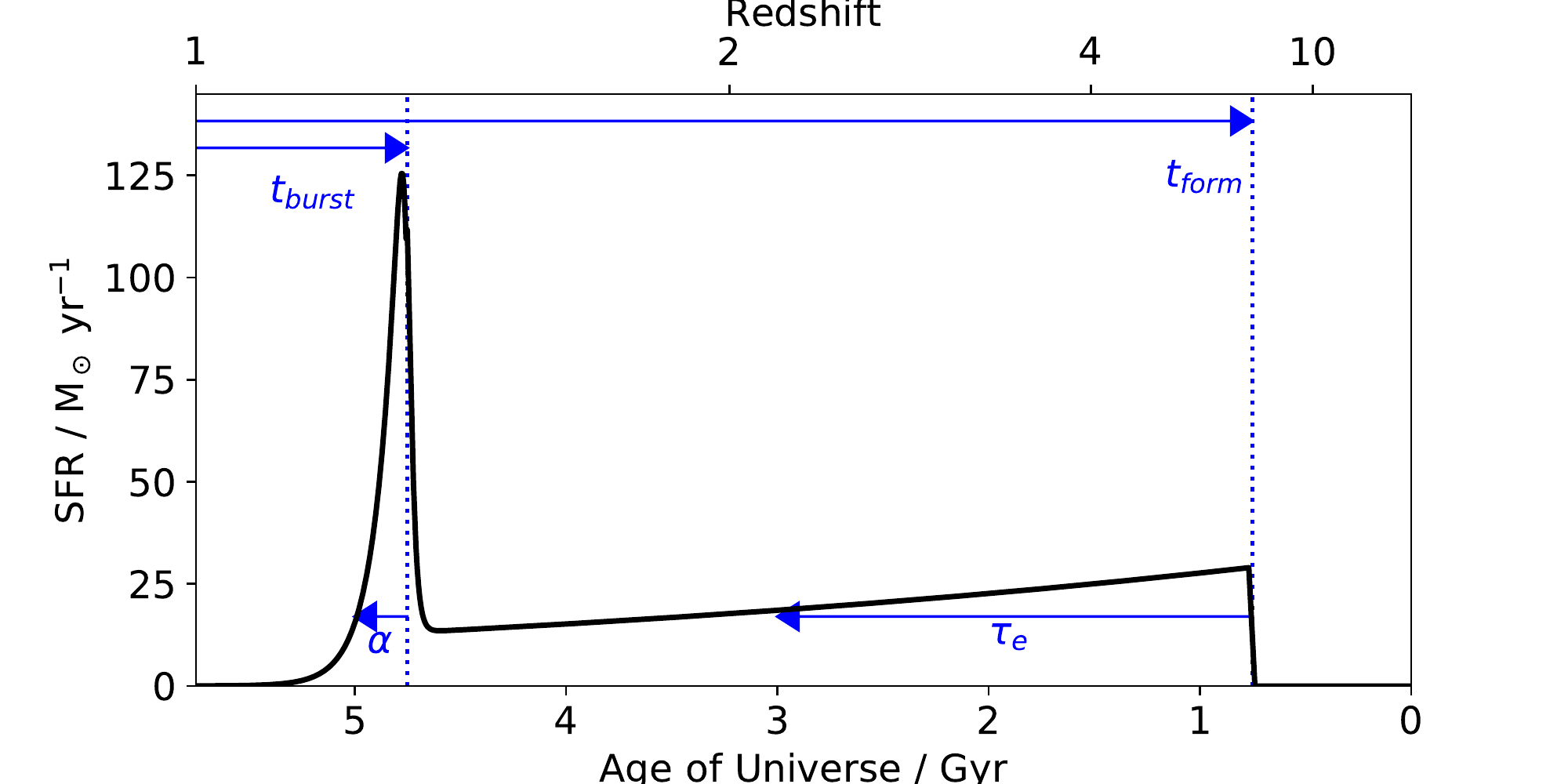}
    \caption{Example model star formation history fitted to the post-starburst galaxies (see Eqn. \ref{eqn:sfh}). In this example, the galaxy has a redshift of 1 which corresponds to a Universe age of 5.75\,Gyr. It has formed $10^{11}$M$_\odot$ of stars and has a burst mass fraction of 10\%, a burst age of $t_{burst}=$1\,Gyr, an age of formation of $t_{form}=5$\,Gyr, an exponential decline time of $\tau_e=5$\,Gyr, and $\alpha=50$ controlling the decline rate of the double powerlaw burst. }
    \label{fig:sfh}
\end{figure}

\begin{figure*}
    \centering
    \includegraphics[width=\textwidth]{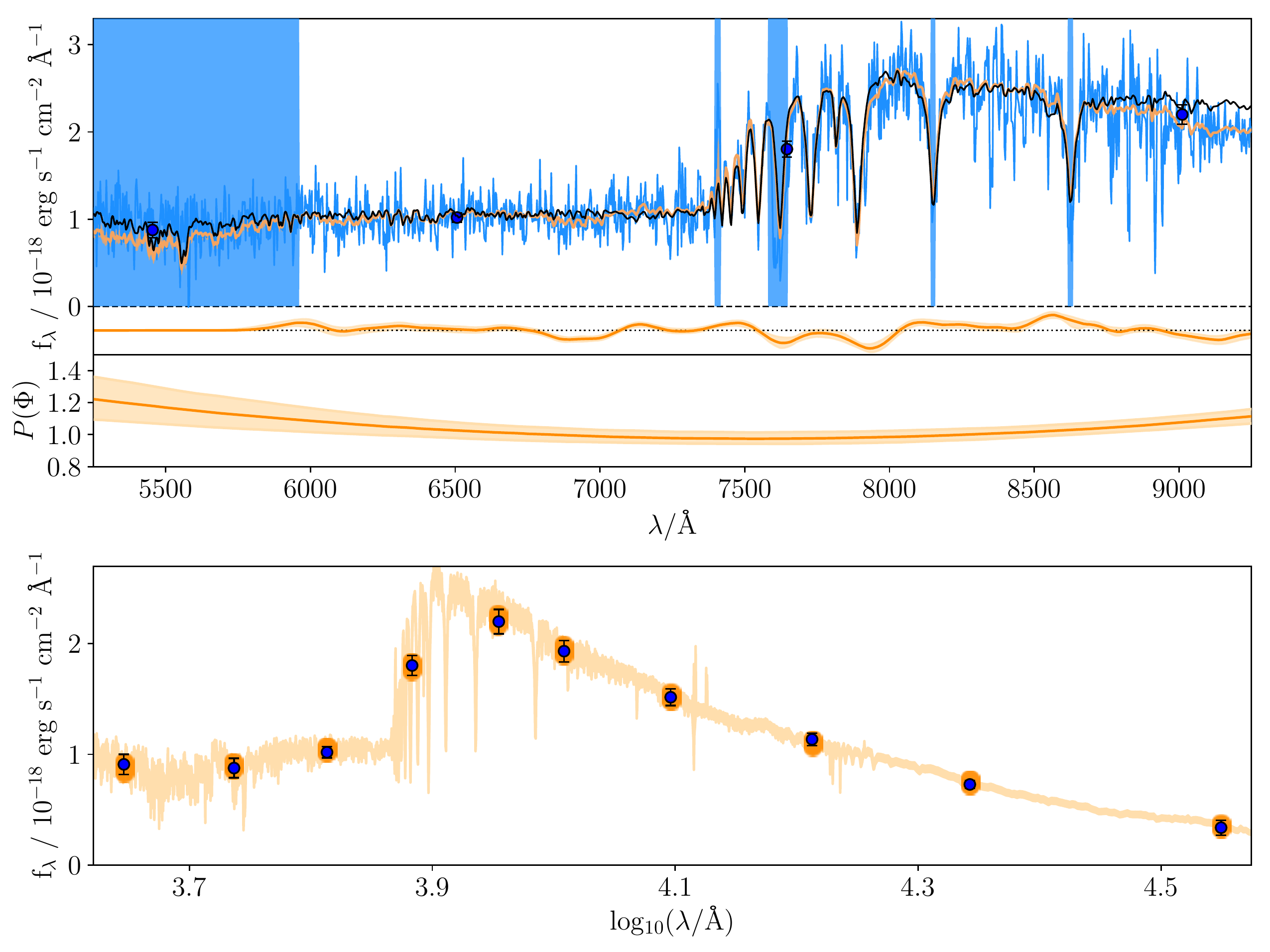}
    \caption{Example {\sc bagpipes} fit to a post-starburst galaxy (DR11 ID 122200, $z=0.99$) using the two component burst model (Eqn.~\ref{eqn:sfh}, Fig.~\ref{fig:sfh}). {\it Top:} the input spectrum in the observed frame (blue) and posterior fit (orange). The fitted model without the calibration corrections is shown in black and the photometry as blue points. The Gaussian process noise component is shown arbitrarily offset below zero for clarity, with the dotted line marking zero correction. The lower panel shows the polynomial spectrophotometric correction. Regions that are masked during fitting are indicated by blue shading. {\it Bottom:} the input photometry (blue), posterior fitted photometry (orange circles) and posterior SED (orange line). The resulting star formation history for this galaxy can be seen in Fig.\ref{fig:sfh_SCpsb}.}
    \label{fig:egfit}
\end{figure*}

The double power-law model used in the initial fits has been shown to work well for quiescent galaxies \citep{CarnallVANDELS2019}. However, it only allows a single rise and fall in star formation over time. For our post-starburst samples, this forces 100\% of the mass into the recent starburst, which may not be realistic.  Following selection of our post-starburst and quiescent samples (described below), we fit a second model to these two samples that allows for a secondary burst of star formation. We choose an old, exponentially declining component with uniform prior on the decline time ($\tau_{e}$) between 300\,Myr and 10\,Gyr and age ($t_{\rm form}$)  between 4\,Gyr and the age of the Universe at the redshift of the galaxy, truncated at the time of the burst. To this we add a double-powerlaw young starburst occurring at a burst time ($t_{burst}$) within the last 2\,Gyr, with a fixed rising slope of $\beta=250$ and variable declining slope ($\alpha$). A variable $\beta$ was attempted, but was poorly constrained by the data and was removed for speed. Varying $\beta$ within reasonable values had no impact on the results, so long as the rise was sufficiently rapid to form a well defined burst. The relative strengths of the two components is characterised by a flat prior on the burst mass fraction ($f_{burst} = M_{burst}/M_{tot}$ where $M_{burst}$ is the integrated mass of stars formed in the burst and $M_{tot}$ is the total integrated mass of stars formed), which can range between 0 and 1. The star formation rate as a function of time $\psi(t)$ is shown in Fig.~\ref{fig:sfh} and can be written as:
\begin{eqnarray}\label{eqn:sfh}
    \psi(t) &\propto& \left. \frac{1-f_{burst}}{\int \psi_{\rm e} dt} 
    \times \psi_e(t) \right|_{t_{\rm form}>t>t_{burst}}\\
    &+& \frac{f_{burst}}{\int \psi_{burst} dt} 
    \times \psi_{burst}(t)
\end{eqnarray}
where the two independent components are given by 
\begin{eqnarray}
    \psi_e(t) &=& \exp^\frac{-t}{\tau_{e}}\\
    \psi_{burst}(t) &=& \left[ \left(\frac{t}{t_{burst}}\right)^\alpha + \left(\frac{t}{t_{burst}}\right)^{-\beta} \right]^{-1}.
\end{eqnarray}
In Appendix \ref{App:dblplaw} we show some examples of fitted SFHs to the post-starburst galaxies assuming the double powerlaw vs. burst model priors. 

\begin{figure*}
    \centering
    \includegraphics[width=\textwidth]{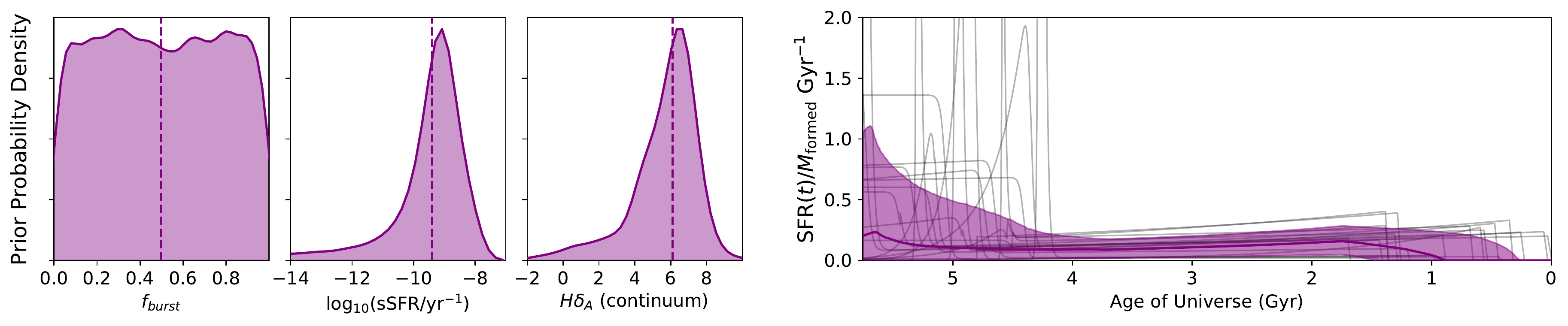}
    \caption{The prior distribution on the burst mass fraction (left, flat by design), sSFR (centre left) and spectral index \hda\ (centre right) for the two component burst model (Eqn.~\ref{eqn:sfh}, Fig.~\ref{fig:sfh}). The dashed line shows the median value. On the right we show the prior median SFH (purple line) with the 16th and 84th percentiles (shaded area), normalised by the mass formed at that time. 20 random draws from the prior SFH are overplotted in grey.  }
    \label{fig:prior}
\end{figure*}

Fig.~\ref{fig:egfit} shows an example {\sc bagpipes} fit to one of the post-starburst galaxies using the two-component model.  The double powerlaw starburst has the advantage over an exponential (tau) starburst of allowing the star formation rate to decay more completely to zero, which was found to be necessary to fit the UV flux of our post-starburst data. Framing the model in terms of burst mass fraction ($f_{burst}$), rather than mass ratio, was required to ensure the prior on the burst mass fraction was flat. A flat prior on mass ratio instead imposes a double horned prior on the burst mass fraction, with strong preferences for either 0 or 1, which resulted in overestimated burst mass fractions for the post-starbursts. Due to the fixed rising slope on $\beta$, any width to the burst is  controlled by $\alpha$ alone. 

Following \citet{CarnallLeja2019}, Fig.~\ref{fig:prior} shows some relevant quantities from the assumed prior SFH distribution, created from 10000 random draws from the model. While the prior distribution for $f_{burst}$ is flat by construction, as with many parameterised star formation history models there is an implicit tight prior on the sSFR. In the centre right panel we show how this in turn leads to a tight prior on the measured \hda, which we will return to below. On the right we show the resulting prior on the star formation rate, normalised by the mass formed by that time (i.e. closely related to sSFR but not accounting for mass loss). At ages $>2$\,Gyr the prior median SFH is a smooth decline, at younger ages this rises allowing for significant amounts of mass to be formed at recent times. 

We additionally altered the dust law slightly compared to the initial fits, allowing the amount of attenuation caused by the birthclouds surrounding stars younger than $10^7$ years to be larger than that in the ISM by a factor $\eta$, using a Gaussian prior with mean of $\eta=3$ and width of 1. This is a crucial part of the modelling of post-starburst galaxies, as a possible source of contamination of the spectroscopic samples may be dust-obscured starbursts, where a high $\eta$ causes a large fraction of light from OB stars to be hidden behind dense dust clouds, leading to the strong Balmer absorption lines from AF stars used to identify post-starburst galaxies. We fixed the slope on the ISM and birthcloud components to 0.7 and 1.3 respectively, following \citet{Wild2007}. 

It is worth noting that burst mass fraction is not a particularly well defined quantity, even for parametric star formation histories. There is nothing to stop the code from replacing the end of the exponential decay with a double power-law ``burst'' in order to extend the star formation history to times within the last 2\,Gyr. The fitted value of $\alpha$ allows a star formation history that is either continuous from the end of the exponential, or increases, or efficiently shuts off. The relative fraction of mass in the double power-law component will depend on how long the star formation needs to run for before shutting off, and can not always be physically interpreted as a burst mass fraction. We therefore additionally calculate the fraction of mass formed within the last 1 and 1.5\,Gyr, which are quantities robust to the form of the fitted star formation history, and as we shall see, usefully separate post-starburst galaxies from others.

\subsection{Sample selection}
Starting from our combined catalogue of all 1439 spectroscopic observations in the UDS, we select only those objects with $0.5<z<1.3$ and a SNR$>3$ in the H$\delta\lambda4100$ spectral region. Below $z=0.5$ the super-colour measurements are less reliable due to increasing rest-frame wavelength separation between the bands and therefore weaker constraints on the sharpness of the Balmer break. Above $z=1.3$ we loose \hda\ from the red end of the spectra. We further remove 3 clear broad line AGN. This gives a starting sample of 694 spectra. The majority (617) are from the UDSz survey, 45 from VANDELS, and 32 from DM16.

Once the initial {\sc bagpipes} fits have been performed, we remove from our sample galaxies where the initial redshift and {\sc bagpipes} redshift differ by more than $\pm0.005$. This indicates that {\sc bagpipes} has been unable to find the correct redshift in the noisy stellar continuum spectra, without the additional information afforded by the masked emission lines. We make this cut to ensure that the spectra add value to the photometry: where {\sc bagpipes} can not find even a redshift from the stellar continuum, it is certain that any spectral indices that we measure from the fit are driven by the photometry alone. This reduces our initial sample to 668. We further remove galaxies for which we are unable to measure one of the spectral indices, leaving 635 galaxies. These final cuts only remove UDSz spectra from the sample, leaving a final sample of 558 from the UDSz survey, 45 from VANDELS, and 32 from DM16.

\begin{table*}
    \caption{Measured properties of both the spectroscopically and photometrically selected post-starburst galaxies. Values given are median values of the posterior distribution, with errors calculated from the 16th and 84th percentiles. Columns are (1) DR11 ID number, (2) redshift, (3) stellar mass as log(M$^*$/M$_\odot$), (4) first super-colour amplitude, (5) second super-colour amplitude, (6) class as defined by super-colours, (7) \hda/\AA\ measured from the fitted continuum model, an asterisk indicates the \hda\ measurement is made from an extrapolated spectrum (see text), (8) W[OII]/\AA (positive in emission), (9) \dn\ measured from the fitted continuum model (statistical errors are small), (10) survey from which the spectrum was taken (DM = PI observations from \citep{Maltby2016}), (11) instrument used to obtain the spectrum.  }
    \centering
    \begin{tabular}{lllllllllll}
\hline
 DR11 ID   & z      & logM*                 & SC1     & SC2    & SC class   & \hda$_{\rm cont}$     & W[OII]                & \dn    & survey   & instrument   \\
\hline
 $33044$   & $0.69$ & $9.86^{+0.1}_{-0.1}$  & $-7.0$  & $10.3$ & PSB        & $8.63^{+0.6}_{-0.5}$  & $6.88^{+1.4}_{-1.2}$  & $1.36$ & UDSz     & VIMOS        \\
 $58883$   & $1.26$ & $11.44^{+0.2}_{-0.1}$ & $-12.9$ & $2.7$  & SF         & *$6.08^{+0.3}_{-0.2}$ & $4.92^{+0.7}_{-0.6}$  & $1.26$ & VANDELS  & VIMOS        \\ 
 $107102$  & $0.54$ & $10.39^{+0.1}_{-0.1}$ & $-21.0$ & $1.2$  & Q          & $5.52^{+0.4}_{-0.5}$  & $4.37^{+0.9}_{-1.1}$  & $1.47$ & UDSz     & VIMOS        \\
 $108365$  & $1.12$ & $10.84^{+0.1}_{-0.1}$ & $-2.9$  & $12.6$ & PSB        & $7.60^{+0.5}_{-0.5}$  & $0.51^{+0.1}_{-0.1}$  & $1.42$ & DM       & VIMOS        \\
 $109922$  & $1.04$ & $11.73^{+0.3}_{-0.4}$ & $-27.6$ & $-9.2$ & D          & $6.31^{+0.7}_{-0.4}$  & $4.74^{+1.8}_{-1.4}$  & $1.29$ & UDSz     & FORS2        \\
 $122200$  & $0.99$ & $10.22^{+0.1}_{-0.1}$ & $15.1$  & $18.4$ & PSB        & $9.89^{+0.3}_{-0.5}$  & $5.01^{+0.3}_{-0.2}$  & $1.25$ & DM       & VIMOS        \\
 $125246$  & $1.28$ & $10.54^{+0.2}_{-0.1}$ & $-8.5$  & $6.0$  & SF         & *$8.04^{+0.6}_{-0.6}$ & $18.19^{+0.7}_{-1.6}$ & $1.33$ & VANDELS  & VIMOS        \\
 $132150$  & $1.14$ & $11.19^{+0.2}_{-0.2}$ & $-17.9$ & $2.0$  & Q          & $5.50^{+0.5}_{-0.5}$  & $4.70^{+0.5}_{-0.3}$  & $1.46$ & DM       & VIMOS        \\
 $133987$  & $1.01$ & $10.94^{+0.1}_{-0.1}$ & $5.6$   & $15.9$ & PSB        & $8.28^{+0.5}_{-0.5}$  & $1.01^{+0.9}_{-0.5}$  & $1.33$ & UDSz     & VIMOS        \\
 $152227$  & $1.15$ & $10.65^{+0.2}_{-0.1}$ & $7.8$   & $14.4$ & PSB        & *$8.37^{+0.6}_{-0.7}$ & $2.91^{+1.1}_{-0.8}$  & $1.32$ & UDSz     & VIMOS        \\
 $153020$  & $1.00$ & $10.75^{+0.1}_{-0.1}$ & $-7.9$  & $11.0$ & PSB        & $7.12^{+0.5}_{-0.7}$  & $4.29^{+0.2}_{-0.5}$  & $1.44$ & DM       & VIMOS        \\
 $153502$  & $1.27$ & $10.84^{+0.2}_{-0.1}$ & $13.0$  & $17.2$ & PSB        & *$9.86^{+0.2}_{-0.3}$ & $4.95^{+0.4}_{-0.3}$  & $1.23$ & DM       & VIMOS        \\
 $157397$  & $1.06$ & $11.07^{+0.3}_{-0.2}$ & $4.8$   & $16.8$ & PSB        & $7.18^{+0.8}_{-0.9}$  & $1.93^{+1.3}_{-0.9}$  & $1.26$ & UDSz     & VIMOS        \\
 $164912$  & $0.62$ & $10.83^{+0.2}_{-0.2}$ & $-10.9$ & $0.5$  & SF         & $5.52^{+0.3}_{-0.5}$  & $0.06^{+0.7}_{-0.5}$  & $1.31$ & UDSz     & VIMOS        \\
 $173705$  & $1.23$ & $10.73^{+0.2}_{-0.1}$ & $-6.2$  & $-1.0$ & SF         & *$5.46^{+0.5}_{-0.6}$ & $4.61^{+1.0}_{-0.3}$  & $1.32$ & UDSz     & FORS2        \\
 $176778$  & $0.51$ & $10.99^{+0.1}_{-0.1}$ & $-26.1$ & $-7.4$ & SF         & $5.63^{+0.3}_{-0.2}$  & $4.74^{+0.7}_{-0.7}$  & $1.34$ & UDSz     & VIMOS        \\
 $182104$  & $0.92$ & $10.38^{+0.1}_{-0.1}$ & $-6.7$  & $10.8$ & PSB        & $5.90^{+0.5}_{-0.6}$  & $2.60^{+0.3}_{-0.4}$  & $1.44$ & DM       & VIMOS        \\
 $186754$  & $1.10$ & $10.52^{+0.2}_{-0.1}$ & $-2.9$  & $12.3$ & PSB        & $8.68^{+0.6}_{-0.5}$  & $4.16^{+0.3}_{-0.2}$  & $1.38$ & DM       & VIMOS        \\
 $187658$  & $1.10$ & $11.03^{+0.2}_{-0.1}$ & $-21.3$ & $-2.5$ & SF         & $8.51^{+0.4}_{-0.5}$  & $20.13^{+1.4}_{-5.3}$ & $1.32$ & DM       & VIMOS        \\
 $187798$  & $0.88$ & $10.29^{+0.2}_{-0.1}$ & $-8.1$  & $10.9$ & PSB        & $5.51^{+1.1}_{-1.2}$  & $0.74^{+0.3}_{-0.3}$  & $1.42$ & UDSz     & VIMOS        \\
 $191179$  & $1.19$ & $11.09^{+0.3}_{-0.2}$ & $-23.7$ & $-2.8$ & Q          & $5.82^{+0.3}_{-0.5}$  & $4.14^{+1.3}_{-0.9}$  & $1.33$ & UDSz     & FORS2        \\
 $193971$  & $0.99$ & $10.49^{+0.2}_{-0.2}$ & $-7.7$  & $-0.0$ & SF         & $5.89^{+0.3}_{-0.4}$  & $1.98^{+3.7}_{-1.7}$  & $1.29$ & UDSz     & FORS2        \\
 $213260$  & $1.17$ & $10.86^{+0.3}_{-0.2}$ & $5.8$   & $17.0$ & PSB        & $8.34^{+0.3}_{-0.5}$  & $2.58^{+0.5}_{-0.6}$  & $1.26$ & UDSz     & VIMOS        \\
 $229763$  & $1.30$ & $11.12^{+0.2}_{-0.1}$ & $-10.5$ & $7.0$  & PSB        & *$5.31^{+0.5}_{-0.7}$ & $1.92^{+0.4}_{-0.3}$  & $1.44$ & UDSz     & FORS2        \\
 $255581$  & $0.64$ & $10.55^{+0.2}_{-0.2}$ & $-21.8$ & $-0.2$ & Q          & $6.00^{+0.7}_{-0.7}$  & $3.02^{+1.0}_{-0.8}$  & $1.45$ & UDSz     & VIMOS        \\

 \hline

 $65602$   & $1.11$ & $10.88^{+0.1}_{-0.1}$ & $-6.7$  & $10.3$ & PSB        & $6.21^{+0.3}_{-0.2}$  & $7.47^{+0.5}_{-0.4}$  & $1.45$ & VANDELS  & VIMOS        \\
 $96779$   & $1.12$ & $10.24^{+0.1}_{-0.1}$ & $-8.0$  & $9.4$  & PSB        & $3.83^{+0.9}_{-0.8}$  & $0.95^{+0.7}_{-0.7}$  & $1.47$ & UDSz     & FORS2        \\
 $98596$   & $1.26$ & $10.58^{+0.2}_{-0.2}$ & $-9.5$  & $10.2$ & PSB        & *$4.03^{+0.6}_{-0.8}$ & $-0.39^{+2.7}_{-1.6}$ & $1.33$ & UDSz     & FORS2        \\
 $108587$  & $1.20$ & $11.04^{+0.1}_{-0.1}$ & $-12.8$ & $8.0$  & PSB        & $3.07^{+0.4}_{-0.6}$  & $-0.18^{+0.1}_{-0.1}$ & $1.54$ & UDSz     & FORS2        \\
 $114152$  & $1.27$ & $10.73^{+0.1}_{-0.1}$ & $-6.2$  & $13.4$ & PSB        & *$3.91^{+0.5}_{-0.6}$ & $2.21^{+0.9}_{-0.7}$  & $1.36$ & UDSz     & FORS2        \\
 $116031$  & $1.28$ & $10.94^{+0.1}_{-0.1}$ & $-12.3$ & $9.3$  & PSB        & *$5.90^{+0.4}_{-0.8}$ & $5.18^{+0.2}_{-0.2}$  & $1.49$ & UDSz     & FORS2        \\
 $124662$  & $1.11$ & $10.18^{+0.2}_{-0.1}$ & $-1.2$  & $11.5$ & PSB        & $5.90^{+0.8}_{-0.6}$  & $7.83^{+0.3}_{-0.4}$  & $1.31$ & VANDELS  & VIMOS        \\
 $125588$  & $1.03$ & $10.34^{+0.1}_{-0.1}$ & $-9.2$  & $8.9$  & PSB        & $4.79^{+0.8}_{-0.9}$  & $-0.16^{+0.4}_{-0.3}$ & $1.45$ & DM       & VIMOS        \\
 $136729$  & $0.57$ & $9.85^{+0.1}_{-0.1}$  & $-8.7$  & $7.6$  & PSB        & $4.71^{+0.7}_{-0.8}$  & $1.90^{+0.9}_{-0.8}$  & $1.47$ & DM       & VIMOS        \\
 $138600$  & $1.27$ & $10.85^{+0.2}_{-0.1}$ & $-10.2$ & $9.2$  & PSB        & *$4.89^{+1.1}_{-1.0}$ & $11.57^{+2.4}_{-2.2}$ & $1.46$ & DM       & VIMOS        \\
 $162358$  & $0.54$ & $9.96^{+0.1}_{-0.1}$  & $-8.3$  & $9.5$  & PSB        & $4.62^{+0.6}_{-0.7}$  & $2.79^{+0.3}_{-0.2}$  & $1.48$ & DM       & VIMOS        \\
 $165790$  & $1.19$ & $11.03^{+0.2}_{-0.2}$ & $-10.3$ & $6.8$  & PSB        & $4.76^{+0.5}_{-0.5}$  & $5.10^{+0.9}_{-1.0}$  & $1.26$ & DM       & VIMOS        \\
 $212238$  & $1.17$ & $10.64^{+0.1}_{-0.1}$ & $-8.7$  & $9.9$  & PSB        & $3.95^{+0.9}_{-0.9}$  & $0.28^{+0.5}_{-0.4}$  & $1.50$ & UDSz     & FORS2        \\
 $240201$  & $0.99$ & $10.96^{+0.2}_{-0.2}$ & $-11.8$ & $7.9$  & PSB        & $4.13^{+0.4}_{-0.5}$  & $0.81^{+0.6}_{-0.8}$  & $1.32$ & UDSz     & VIMOS        \\
\hline
\end{tabular}
    \label{tab:sample}
\end{table*}

\begin{figure*}
    \centering
    \includegraphics[width=\textwidth]{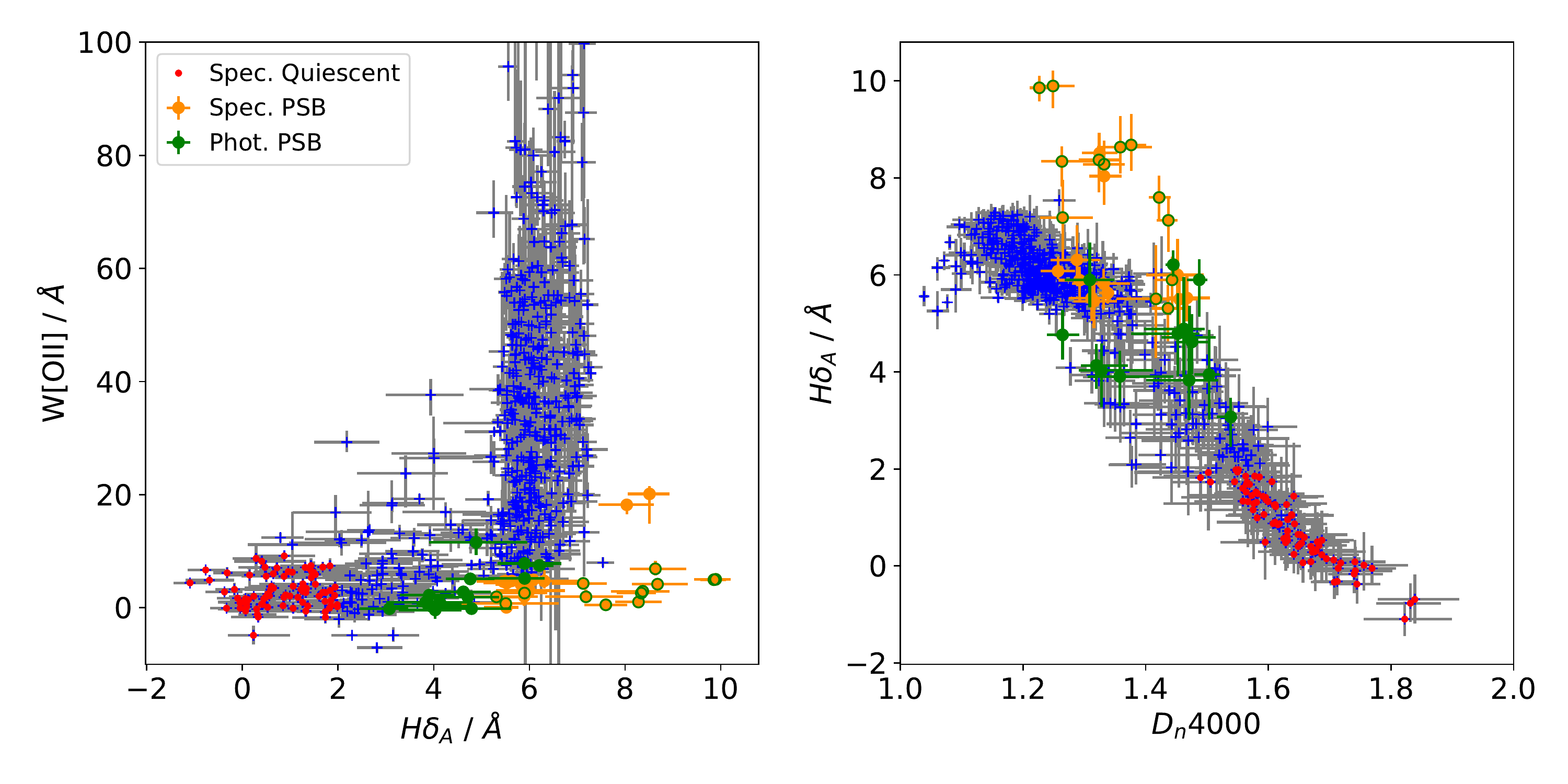}
    \caption{{\it Left:} Equivalent widths of the stellar continuum absorption line \hda\ (positive in absorption) vs. [OII] nebular emission line (positive in emission) for our full sample of spectroscopically observed galaxies.  {\it Right:} \dn\ vs. \hda\ for the same sample. The spectroscopic samples of post-starburst and quiescent galaxies studied in this paper are marked as orange and red points respectively. The photometrically identified post-starbursts are marked as green points. Post-starbursts that are identified both photometrically and spectroscopically are marked as orange with green outer rings. The remaining galaxies not used in this paper are marked as blue crosses. \hda\ and \dn\ are both measured from the continuum fit to the photometry and spectroscopy, to improve signal as well as remove infilling caused by emission lines. The same models are used to define the continuum from which W[OII] is measured.}
    \label{fig:sample}
\end{figure*}

\begin{figure}
    \centering
    \includegraphics[width=\columnwidth]{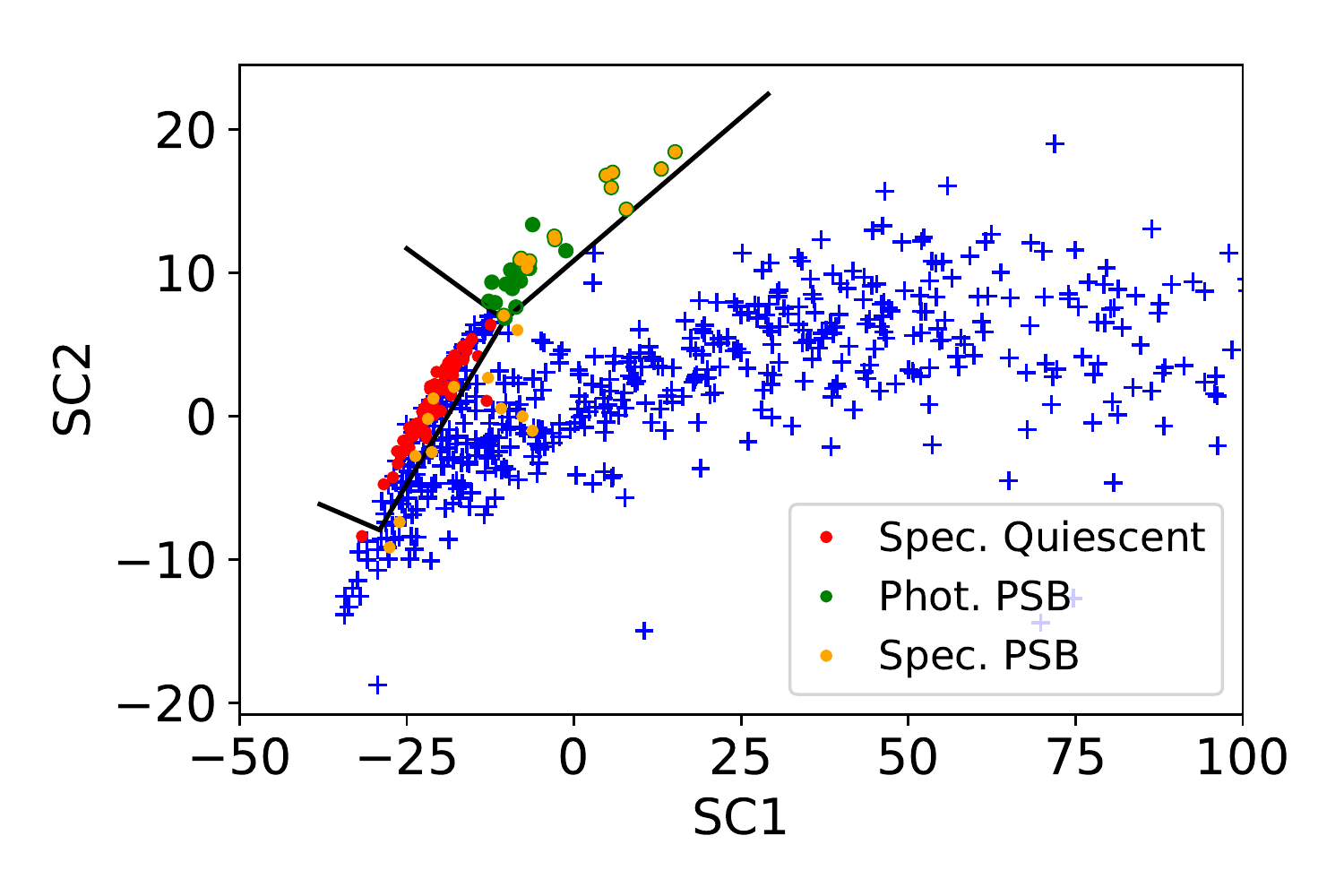}
    \caption{Principal component amplitudes (``super-colours'') SC1 vs. SC2 for our sample of spectroscopically observed galaxies.  The first super-colour amplitude (SC1) describes the overall red/blue colour of the SED, while the second (SC2) relates to the strength of the 4000\AA\ or Balmer break. Symbols are as in Fig. \ref{fig:sample}. The black demarcation lines indicate the main super-colour class boundaries:  quiescent to the left, star-forming to the right, post-starbursts at high SC2 and dusty star-forming galaxies at low SC2.   }
    \label{fig:SC}
\end{figure}

\begin{figure*}
    \centering
    \includegraphics[width=\columnwidth]{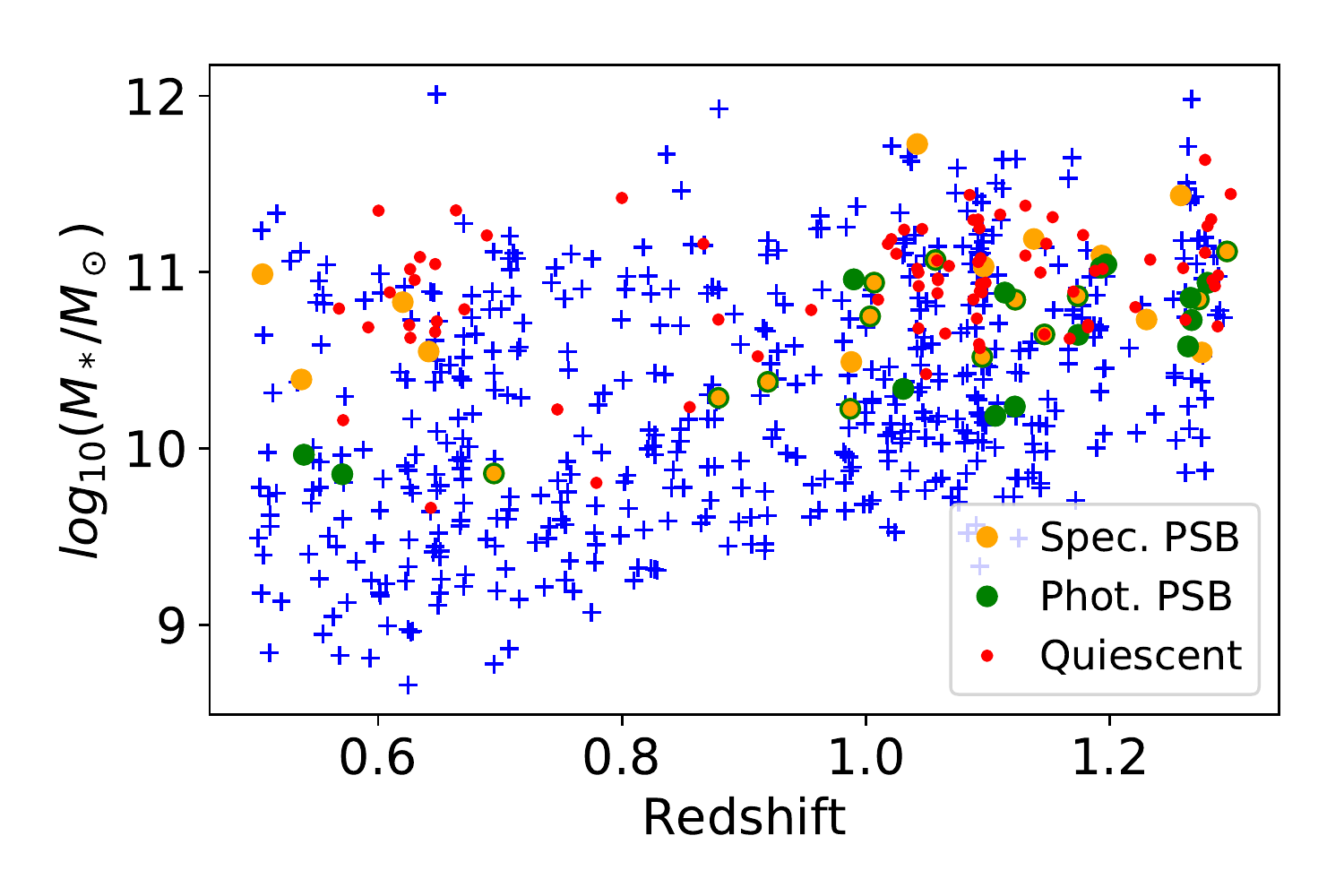}
    \includegraphics[width=\columnwidth]{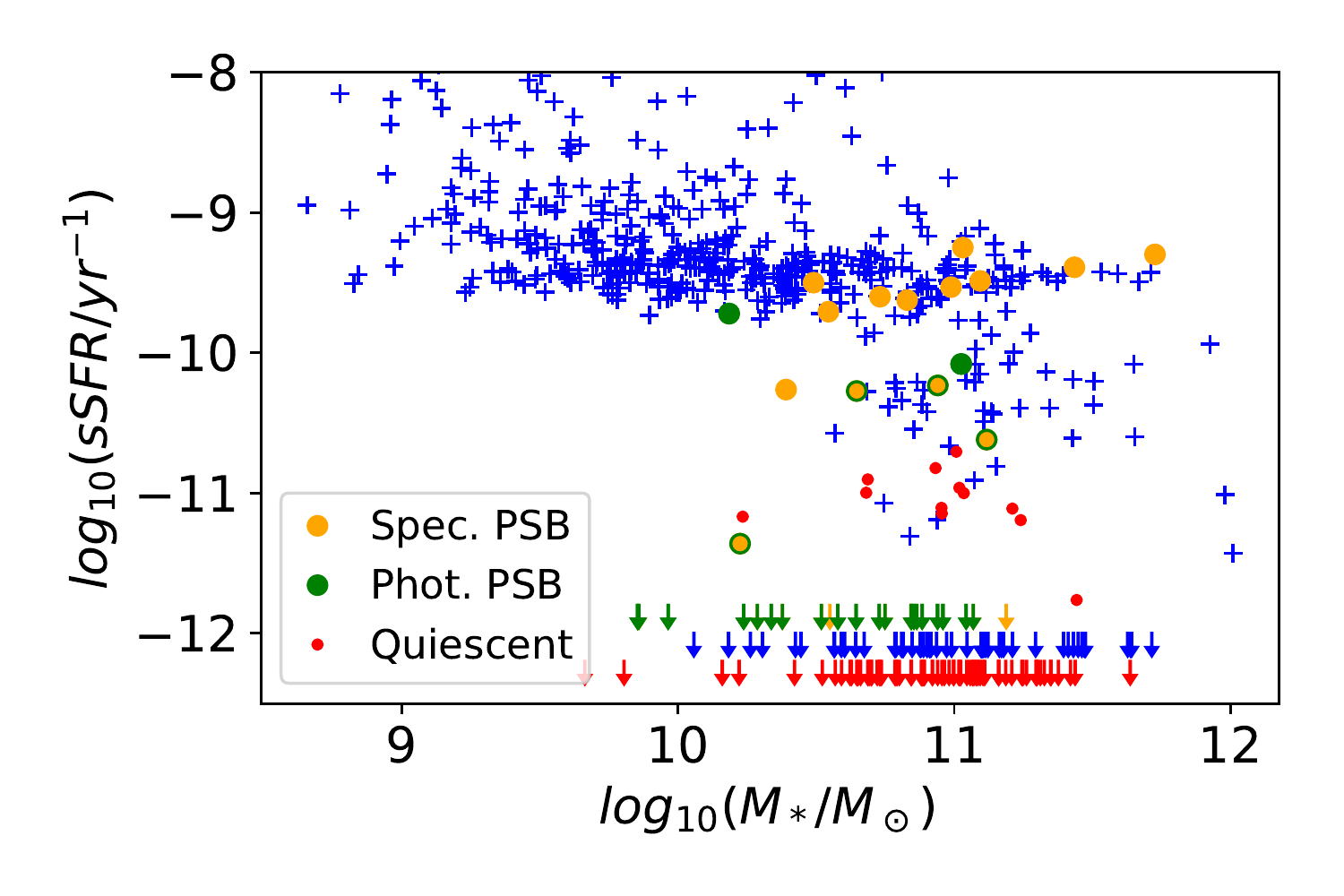}
    \caption{{\it Left:} The redshift vs. stellar mass distribution of our sample of spectroscopically observed galaxies with $0.5<z<1.3$. {\it Right:} The stellar mass vs. sSFR distribution, where both mass and sSFR are the median value from the posterior distribution of fitted double power law models. Upper limits are given where sSFR$<10^{-12}$/yr as values below this are dependent entirely on the prior SFH assumed. For clarity an arbitrary offset is applied between upper limits for different samples. Symbols are as in Fig.~\ref{fig:sample}.}
    \label{fig:massz}
\end{figure*}

We identify spectroscopic post-starbursts with  \hda$>5$\AA\ and W[OII]$<5$\AA, where \hda\ and the continuum for W[OII] is measured from the fitted double powerlaw model. Due the difference between measurements of the equivalent width of H$\delta$ this is a slightly more conservative cut than used in DM16. We additionally allow two objects with \hda$>8$\AA, even though they have W[OII]$\sim20$\AA. These two objects lie significantly above the main sequence in \hda\ and the [OII] emission may arise from an AGN. We shift slightly the quiescent limits compared to DM16, to allow for larger errors on W[OII] and to identify truly quiescent galaxies with \hda$<2$\AA\ and W[OII]$<10$\AA. We use the convention that \hda\ is positive in absorption, while W[OII] is positive in emission. This selects 27 post-starburst galaxies and 114 quiescent galaxies. We carefully inspect each of the post-starburst galaxy spectra and fits, and remove a further 2 spurious objects, leaving a total sample of 25. 

We additionally identify 14 objects that are classified as post-starburst galaxies photometrically using the super-colours (see Section \ref{sec:data}), but are not included in our spectroscopic classification of post-starburst galaxies, giving a total sample of 39. Post-starburst galaxies are identified photometrically using the first and second super-colours, which efficiently separates intermediate age galaxies with a well defined Balmer break, weak UV continuum and continually declining NIR SED, from those with older and younger stellar populations \citep{Wild2016}. Table \ref{tab:sample} lists the main measurements of the selected post-starburst galaxies, with those selected photometrically listed at the bottom. 

For 10 of our post-starbursts the redder side of the \hda\ index falls into a part of the spectrum that has poor spectral calibration and has therefore been masked during the fit. We still measure \hda\ from the fitted model, which naturally extends beyond the range of the fitted data, as the remaining Balmer lines constrain the strength of the H$\delta$ line well. Indeed, the errors on these measurements are typical for the sample as a whole. These extrapolated measurements are marked by an asterisk in the Table. We note that the cut on W[OII] in our spectroscopic sample will exclude post-starburst galaxies that had a more recent burst, or are quenching more slowly, as well as galaxies with strong narrow line AGN and/or shock ionisation \citep{Yan2006,Wild2007,Wild2010,Alatalo2016}. In the case of narrow line AGN or shock ionisation the photometric selection should include them, however the sample will likely be incomplete at  younger ages and for slightly longer duration bursts \citep{Wild2007}. 

The left panel of Fig.~\ref{fig:sample} shows the selection of our spectroscopic post-starburst (orange) and quiescent (red) subsamples, using the W[OII] and \hda\ line measurements from the complete spectroscopic measurements (blue crosses). The green points show the position of the photometrically selected post-starburst galaxies, and orange points ringed with green are selected by both methods. We see that the majority of the post-starbursts selected purely photometrically have slightly lower \hda\ than the limit imposed on the spectroscopic selection, while a few have stronger W[OII], which may arise from AGN emission or simply errors on our W[OII] measurements. The right panel of Fig.~\ref{fig:sample} shows the continuum \dn\ vs. \hda\ measurements for our sample. We see that a large fraction of the spectroscopic post-starburst galaxies lie above the main sequence of star-forming galaxies, however, the additional cut on W[OII] potentially allows us to identify galaxies within the main sequence that have prematurely shut down their star formation. Those post-starbursts selected purely photometrically lie within the ``green valley'' with lower \hda\ and stronger \dn\ than their spectroscopically selected counterparts, perhaps indicating older burst ages. We see a strong correspondence between the \hda\ vs. W[OII] selection and \dn\ vs. \hda\ measurements for the quiescent galaxies, with the majority of the galaxies with high \dn\ selected as quiescent. 

Fig.~\ref{fig:SC} shows the position of our spectroscopically observed galaxies on the super-colour diagram (see Section \ref{sec:data}). The spectroscopically selected post-starbursts and quiescent galaxies are marked as orange and red points respectively, while the PSBs selected by super-colours alone are marked as green points. The remaining sample of spectroscopic observations are shown as blue crosses. 13/25 spectroscopic post-starbursts lie in the post-starburst region in SC space, with the remainder lying in the quiescent (4), dusty (1) or low sSFR star-forming population (7).  While at first glance this is not a surprise -- we expect spectroscopy to be far more sensitive at identifying weaker and older post-starburst galaxies than photometry -- we will use the derived star formation histories to confirm the reasons for these differences in classifications below. The vast majority of the spectroscopically selected quiescent galaxies lie in the super-colour red-sequence (81/83).

Returning briefly to Fig.~\ref{fig:sample}, we note that we measure a very narrow range of \hda\ for the star-forming galaxies, that does not appear to change much with W[OII]. As we show in Fig.~\ref{fig:prior} this is an artefact of our fitting procedure. Strict priors on the star formation history, as used in parametric fitting methods, enforce an unphysically tight range in sSFR \citep{CarnallLeja2019}. The small range in continuum \hda\ measured from the models is another manifestation of this effect, meaning that higher quality data is required to measure a value outside of this prior range when the signal is weak compared to if the signal is strong. An alternative would be to use the model to subtract off the emission line flux as is done in e.g. the MPA-JHU SDSS catalogues \citep{Kauffmann2003,Brinchmann2004}, however, we found the quality of the spectra was not sufficient for this to be reliable.  We verified that our selection was independent of the exact details of the prior used, by repeating it on the double powerlaw fit, as well as a fit without Gaussian process noise included. This latter check reassured us that the Gaussian process noise was not being used to alter the equivalent width of the absorption features. 

The left panel of Fig.~\ref{fig:massz} shows the stellar mass vs. redshift distribution of our sample of spectroscopically observed galaxies, assuming the double power-law star formation history. We find stellar masses 0.15\,dex larger on average than those derived in \citet{Wild2016} from fits of exponentially declining models with stochastic bursts to the super-colours. We see that the majority of the quiescent and post-starburst galaxies have stellar mass $10^{10}<$M*/M$_\odot<5\times10^{11}$; this is largely due to completeness limits of these higher mass-to-light ratio galaxies \citep[see][for more details of completeness limits]{Wild2016}. Our sample of spectroscopic post-starbursts include 5 with $z<0.8$, while the remaining 20 have $0.8<z<1.3$. The spectroscopically observed quiescent galaxies have a similar mass-redshift distribution to the post-starbursts, although again we refer the reader to \citet{Wild2016} for completeness corrected stellar mass functions where differences are seen between post-starburst and quiescent galaxies. The right panel of Fig.~\ref{fig:massz} shows the stellar mass vs. sSFR of the sample, where the sSFR is the median value from the fitted posterior double powerlaw star formation histories, with SFR averaged over the last 100\,Myr. This comes from the stellar continuum fit, rather than a direct estimate from the UV, far-infrared or nebular emission lines. It is important to remember that sSFR measured in this way is strongly affected by the assumed priors, in particular the tightness of the star-forming main sequence and the exact values for the quiescent galaxies, and therefore the values should only be used for comparison between samples \citep{CarnallLeja2019}. We set a lower limit of $10^{-12}$/yr as values below this are entirely dependent on the prior SFH assumed and do not reliably indicate a distinction of objects. We see that the quiescent sample have low measured sSFR, as do the majority of the photometrically selected post-starburst. On the other hand, some of the post-starbursts selected on \hda\ and W[OII] alone lie on the blue sequence. 

Finally, we cross-match our PSB sample with the X-ray catalogues of \citet{Ueda2008} and \citet{Kocevski2018}, finding only 1 match (ID 125246). The AGN content of PSBs is being explored in more detail in Almaini et al (in preparation).

\section{Results}\label{sec:results}
\begin{table*}
     \caption{Fitted properties of the spectroscopically and photometrically selected post-starburst galaxies. Values given are median values of the posterior model distribution, with errors calculated from the 16th and 84th percentiles. Columns are (1) DR11 ID number, (2) stellar mass, (3) burst mass fraction (only for burst-like SFHs), (4) burst age, (5) magnitudes of attenuation in the $V$ band, (6) metallicity relative to solar}
    \begin{tabular}{llllll}
\hline
 DR11 ID   & $\log M^*/M_\odot$        & $f_{burst}$           & $t_{burst}$/Gyr        & $A_V$/mag              & $Z/Z_\odot$                 \\
\hline
 $33044$   & $9.93^{+0.05}_{-0.04}$  & $0.75^{+0.2}_{-0.2}$  & $0.50^{+0.1}_{-0.08}$  & $1.18^{+0.2}_{-0.2}$   & $0.68^{+0.4}_{-0.3}$        \\
 $58883$   & $11.37^{+0.04}_{-0.04}$ & --                    & --                     & $1.84^{+0.1}_{-0.1}$   & $0.11^{+0.05}_{-0.04}$      \\
 $107102$  & $10.51^{+0.05}_{-0.06}$ & $0.67^{+0.2}_{-0.3}$  & $1.04^{+0.1}_{-0.1}$   & $1.41^{+0.1}_{-0.1}$   & $0.44^{+0.2}_{-0.1}$        \\
 $108365$  & $10.83^{+0.04}_{-0.05}$ & $0.85^{+0.1}_{-0.2}$  & $0.62^{+0.06}_{-0.07}$ & $0.46^{+0.09}_{-0.1}$  & $1.77^{+0.5}_{-0.4}$        \\
 $109922$  & $11.12^{+0.04}_{-0.04}$ & --                    & --                     & $1.95^{+0.03}_{-0.07}$ & $2.23^{+0.2}_{-0.4}$        \\
 $122200$  & $10.23^{+0.04}_{-0.05}$ & $0.84^{+0.1}_{-0.2}$  & $0.33^{+0.09}_{-0.06}$ & $0.67^{+0.2}_{-0.1}$   & $1.43^{+0.5}_{-0.4}$        \\
 $125246$  & $10.55^{+0.05}_{-0.04}$ & $0.70^{+0.2}_{-0.2}$  & $0.72^{+0.1}_{-0.1}$   & $1.26^{+0.2}_{-0.2}$   & $0.76^{+0.5}_{-0.3}$        \\
 $132150$  & $11.08^{+0.05}_{-0.05}$ & $0.11^{+0.1}_{-0.07}$ & $0.45^{+0.2}_{-0.1}$   & $0.88^{+0.2}_{-0.2}$   & $1.12^{+0.9}_{-0.5}$        \\
 $133987$  & $10.94^{+0.04}_{-0.04}$ & $0.77^{+0.2}_{-0.2}$  & $0.41^{+0.1}_{-0.07}$  & $0.59^{+0.2}_{-0.1}$   & $2.10^{+0.2}_{-0.4}$        \\
 $152227$  & $10.63^{+0.05}_{-0.06}$ & $0.69^{+0.2}_{-0.2}$  & $0.59^{+0.1}_{-0.1}$   & $0.50^{+0.2}_{-0.2}$   & $1.23^{+0.8}_{-0.6}$        \\
 $153020$  & $10.78^{+0.03}_{-0.03}$ & $0.86^{+0.1}_{-0.2}$  & $0.80^{+0.07}_{-0.08}$ & $0.79^{+0.1}_{-0.1}$   & $0.86^{+0.3}_{-0.2}$        \\
 $153502$  & $10.80^{+0.05}_{-0.05}$ & $0.82^{+0.1}_{-0.2}$  & $0.22^{+0.04}_{-0.02}$ & $0.89^{+0.1}_{-0.1}$   & $2.06^{+0.3}_{-0.6}$        \\
 $157397$  & $11.03^{+0.06}_{-0.06}$ & $0.38^{+0.2}_{-0.2}$  & $0.50^{+0.2}_{-0.1}$   & $0.97^{+0.2}_{-0.2}$   & $0.12^{+0.07}_{-0.05}$      \\
 $164912$  & $10.47^{+0.06}_{-0.06}$ & --                    & --                     & $0.85^{+0.3}_{-0.2}$   & $0.85^{+0.6}_{-0.4}$        \\
 $173705$  & $10.62^{+0.04}_{-0.03}$ & --                    & --                     & $0.55^{+0.1}_{-0.1}$   & $1.39^{+0.5}_{-0.4}$        \\
 $176778$  & $11.06^{+0.04}_{-0.04}$ & --                    & --                     & $1.93^{+0.05}_{-0.07}$ & $2.05^{+0.3}_{-0.4}$        \\
 $182104$  & $10.39^{+0.05}_{-0.04}$ & $0.67^{+0.2}_{-0.3}$  & $0.93^{+0.1}_{-0.1}$   & $0.60^{+0.1}_{-0.1}$   & $0.53^{+0.3}_{-0.2}$        \\
 $186754$  & $10.51^{+0.05}_{-0.05}$ & $0.68^{+0.2}_{-0.2}$  & $0.54^{+0.09}_{-0.06}$ & $0.78^{+0.1}_{-0.1}$   & $0.99^{+0.5}_{-0.3}$        \\
 $187658$  & $11.12^{+0.05}_{-0.04}$ & $0.76^{+0.2}_{-0.2}$  & $0.53^{+0.08}_{-0.08}$ & $1.93^{+0.05}_{-0.1}$  & $2.14^{+0.3}_{-0.4}$        \\
 $187798$  & $10.35^{+0.05}_{-0.05}$ & $0.74^{+0.2}_{-0.3}$  & $0.99^{+0.1}_{-0.1}$   & $0.96^{+0.2}_{-0.2}$   & $0.23^{+0.08}_{-0.06}$      \\
 $191179$  & $11.01^{+0.05}_{-0.05}$ & --                    & --                     & $1.76^{+0.1}_{-0.2}$   & $0.32^{+0.2}_{-0.1}$        \\
 $193971$  & $10.22^{+0.04}_{-0.04}$ & --                    & --                     & $0.52^{+0.3}_{-0.2}$   & $1.21^{+0.5}_{-0.4}$        \\
 $213260$  & $10.72^{+0.08}_{-0.08}$ & $0.72^{+0.2}_{-0.2}$  & $0.41^{+0.1}_{-0.07}$  & $1.28^{+0.2}_{-0.2}$   & $0.11^{+0.1}_{-0.06}$       \\
 $229763$  & $11.04^{+0.06}_{-0.07}$ & $0.50^{+0.3}_{-0.3}$  & $1.33^{+0.3}_{-0.4}$   & $0.71^{+0.1}_{-0.2}$   & $0.49^{+0.2}_{-0.1}$        \\
 $255581$  & $10.46^{+0.05}_{-0.06}$ & $0.58^{+0.3}_{-0.2}$  & $0.82^{+0.1}_{-0.1}$   & $1.66^{+0.2}_{-0.2}$   & $0.33^{+0.6}_{-0.09}$ \\\hline
 $65602$   & $10.81^{+0.03}_{-0.03}$ & $0.91^{+0.07}_{-0.1}$ & $1.00^{+0.07}_{-0.09}$ & $0.62^{+0.08}_{-0.08}$ & $0.43^{+0.08}_{-0.04}$      \\
 $96779$   & $10.21^{+0.06}_{-0.09}$ & $0.51^{+0.3}_{-0.3}$  & $1.01^{+0.2}_{-0.2}$   & $0.31^{+0.2}_{-0.2}$   & $0.55^{+0.4}_{-0.2}$        \\
 $98596$   & $10.42^{+0.1}_{-0.09}$  & $0.50^{+0.3}_{-0.3}$  & $0.99^{+0.4}_{-0.2}$   & $0.71^{+0.3}_{-0.4}$   & $0.36^{+0.8}_{-0.3}$        \\
 $108587$  & $11.03^{+0.05}_{-0.04}$ & $0.78^{+0.2}_{-0.2}$  & $1.28^{+0.1}_{-0.1}$   & $0.55^{+0.09}_{-0.1}$  & $0.47^{+0.2}_{-0.07}$       \\
 $114152$  & $10.66^{+0.05}_{-0.08}$ & $0.56^{+0.3}_{-0.3}$  & $1.08^{+0.2}_{-0.1}$   & $0.70^{+0.2}_{-0.2}$   & $0.23^{+0.1}_{-0.1}$        \\
 $116031$  & $11.06^{+0.04}_{-0.04}$ & $0.82^{+0.1}_{-0.2}$  & $1.01^{+0.07}_{-0.08}$ & $1.04^{+0.1}_{-0.1}$   & $0.47^{+0.2}_{-0.1}$        \\
 $124662$  & $10.14^{+0.07}_{-0.06}$ & $0.65^{+0.2}_{-0.3}$  & $1.35^{+0.3}_{-0.3}$   & $0.99^{+0.2}_{-0.2}$   & $0.27^{+0.2}_{-0.1}$        \\
 $125588$  & $10.34^{+0.07}_{-0.08}$ & $0.49^{+0.3}_{-0.2}$  & $0.94^{+0.2}_{-0.2}$   & $0.61^{+0.2}_{-0.3}$   & $0.36^{+0.6}_{-0.2}$        \\
 $136729$  & $9.80^{+0.07}_{-0.07}$  & $0.33^{+0.3}_{-0.2}$  & $0.86^{+0.3}_{-0.3}$   & $0.20^{+0.2}_{-0.1}$   & $1.16^{+0.4}_{-0.3}$        \\
 $138600$  & $10.83^{+0.07}_{-0.07}$ & $0.51^{+0.3}_{-0.3}$  & $0.84^{+0.2}_{-0.2}$   & $0.53^{+0.2}_{-0.2}$   & $0.95^{+0.7}_{-0.6}$        \\
 $162358$  & $9.95^{+0.05}_{-0.06}$  & $0.25^{+0.2}_{-0.1}$  & $0.80^{+0.2}_{-0.2}$   & $0.15^{+0.1}_{-0.1}$   & $1.09^{+0.3}_{-0.2}$        \\
 $165790$  & $10.88^{+0.07}_{-0.06}$ & --                    & --                     & $1.25^{+0.3}_{-0.2}$   & $0.06^{+0.06}_{-0.04}$      \\
 $212238$  & $10.65^{+0.08}_{-0.06}$ & $0.73^{+0.2}_{-0.2}$  & $1.01^{+0.1}_{-0.1}$   & $0.28^{+0.2}_{-0.1}$   & $0.96^{+0.6}_{-0.4}$        \\
 $240201$  & $10.74^{+0.09}_{-0.09}$ & $0.72^{+0.2}_{-0.3}$  & $1.00^{+0.6}_{-0.2}$   & $0.94^{+0.2}_{-0.3}$   & $0.33^{+0.6}_{-0.3}$        \\
\hline
\end{tabular}
    \label{tab:pipes1}
\end{table*}

\begin{table*}
     \caption{Derived properties of the spectroscopically and photometrically selected post-starburst galaxies calculated from their fitted star formation histories. Values given are median values of the posterior model distribution, with errors calculated from the 16th and 84th percentiles. Columns are (1) DR11 ID number, (2) fraction of mass formed in the last 1\,Gyr, (3) fraction of mass formed in the last 1.5\,Gyr, (4) visibility time spent in super-colour post-starburst selection box, (5) maximum SFR in past 2\,Gyr, (6) time taken to quench from peak to sSFR$=0.2/t_H$ in Gyr, where $t_H$ is the age of the Universe at the redshift of quenching, (7) time taken to quench from sSFR$=1/t_H$ to $0.2/t_H$ in Gyr, (8) 0.15th and 16th percentiles of peak to pre-burst SFR ratio, (9) classification of SFH based on SFR$_{ratio}$ (b=burst; pb=possible burst; sf=star forming).}
      \begin{tabular}{lllllllll}
\hline
 DR11 ID   & $f_{1Gyr}$               & $f_{1.5Gyr}$           & log($t_{vis}$)         & SFR$_{max}$            & $\tau_{q1}$            & $\tau_{q2}$             & SFR$_{ratio}$   & class   \\
\hline
 $33044$   & $0.76^{+0.2}_{-0.2}$     & $0.78^{+0.2}_{-0.2}$   & $8.52^{+0.2}_{-0.3}$   & $124^{+59}_{-42}$      & $0.17^{+0.2}_{-0.08}$  & $0.04^{+0.05}_{-0.02}$  & $>14.6,74.1$    & b       \\
 $58883$   & $0.22^{+0.06}_{-0.05}$   & $0.33^{+0.1}_{-0.07}$  & --                     & $126^{+18}_{-18}$      & --                     & --                      & $>0.7,1.0$      & sf      \\
 $107102$  & $0.27^{+0.2}_{-0.2}$     & $0.69^{+0.2}_{-0.2}$   & --                     & $263^{+330}_{-152}$    & $0.33^{+0.6}_{-0.2}$   & $0.09^{+0.2}_{-0.07}$   & $>3.7,19.2$     & b       \\
 $108365$  & $0.86^{+0.1}_{-0.1}$     & $0.87^{+0.09}_{-0.1}$  & $8.79^{+0.1}_{-0.1}$   & $1706^{+683}_{-624}$   & $0.11^{+0.1}_{-0.05}$  & $0.02^{+0.03}_{-0.01}$  & $>35.2,116.3$   & b       \\
 $109922$  & $0.19^{+0.06}_{-0.04}$   & $0.28^{+0.07}_{-0.05}$ & --                     & $70^{+32}_{-13}$       & --                     & --                      & $>0.7,0.9$      & sf      \\
 $122200$  & $0.85^{+0.1}_{-0.2}$     & $0.86^{+0.1}_{-0.1}$   & $8.67^{+0.1}_{-0.2}$   & $236^{+145}_{-66}$     & $0.22^{+0.2}_{-0.1}$   & $0.05^{+0.04}_{-0.03}$  & $>35.7,102.5$   & b       \\
 $125246$  & $0.71^{+0.2}_{-0.2}$     & $0.74^{+0.2}_{-0.2}$   & --                     & $179^{+33}_{-41}$      & $0.68^{+0.1}_{-0.1}$   & $0.21^{+0.05}_{-0.07}$  & $>4.5,13.6$     & b       \\
 $132150$  & $0.19^{+0.1}_{-0.05}$    & $0.28^{+0.1}_{-0.06}$  & --                     & $340^{+248}_{-185}$    & $0.10^{+0.3}_{-0.06}$  & $0.04^{+0.1}_{-0.03}$   & $>0.5,2.5$      & pb      \\
 $133987$  & $0.79^{+0.1}_{-0.2}$     & $0.81^{+0.1}_{-0.2}$   & $8.60^{+0.2}_{-0.2}$   & $1046^{+786}_{-314}$   & $0.24^{+0.2}_{-0.2}$   & $0.06^{+0.07}_{-0.04}$  & $>16.9,69.0$    & b       \\
 $152227$  & $0.71^{+0.2}_{-0.2}$     & $0.73^{+0.2}_{-0.2}$   & $8.80^{+0.1}_{-0.2}$   & $286^{+86}_{-58}$      & $0.55^{+0.2}_{-0.2}$   & $0.17^{+0.06}_{-0.07}$  & $>7.0,24.3$     & b       \\
 $153020$  & $0.86^{+0.1}_{-0.2}$     & $0.88^{+0.09}_{-0.1}$  & $8.79^{+0.1}_{-0.1}$   & $1235^{+794}_{-600}$   & $0.16^{+0.2}_{-0.09}$  & $0.04^{+0.08}_{-0.02}$  & $>16.4,108.8$   & b       \\
 $153502$  & $0.84^{+0.1}_{-0.1}$     & $0.86^{+0.1}_{-0.1}$   & $8.58^{+0.2}_{-0.1}$   & $1702^{+492}_{-431}$   & $0.08^{+0.07}_{-0.03}$ & $0.02^{+0.02}_{-0.006}$ & $>46.5,118.3$   & b       \\
 $157397$  & $0.42^{+0.2}_{-0.1}$     & $0.47^{+0.2}_{-0.1}$   & $9.05^{+0.07}_{-0.1}$  & $758^{+698}_{-302}$    & $0.18^{+0.4}_{-0.1}$   & $0.06^{+0.2}_{-0.04}$   & $>4.3,12.9$     & b       \\
 $164912$  & $0.10^{+0.04}_{-0.02}$   & $0.16^{+0.06}_{-0.04}$ & --                     & $11^{+3}_{-3}$         & --                     & --                      & $>0.6,0.7$      & sf      \\
 $173705$  & $0.09^{+0.03}_{-0.02}$   & $0.15^{+0.05}_{-0.03}$ & --                     & $19^{+4}_{-4}$         & --                     & --                      & $>0.5,0.6$      & sf      \\
 $176778$  & $0.18^{+0.04}_{-0.03}$   & $0.27^{+0.08}_{-0.05}$ & --                     & $50^{+17}_{-8}$        & --                     & --                      & $>0.7,1.0$      & sf      \\
 $182104$  & $0.49^{+0.3}_{-0.2}$     & $0.70^{+0.2}_{-0.2}$   & $8.90^{+0.03}_{-0.08}$ & $332^{+261}_{-172}$    & $0.20^{+0.3}_{-0.1}$   & $0.05^{+0.1}_{-0.04}$   & $>7.8,31.9$     & b       \\
 $186754$  & $0.71^{+0.2}_{-0.2}$     & $0.74^{+0.2}_{-0.2}$   & $8.79^{+0.1}_{-0.2}$   & $583^{+324}_{-225}$    & $0.13^{+0.2}_{-0.07}$  & $0.03^{+0.06}_{-0.02}$  & $>20.2,56.1$    & b       \\
 $187658$  & $0.78^{+0.1}_{-0.2}$     & $0.80^{+0.1}_{-0.2}$   & --                     & $877^{+141}_{-121}$    & $0.52^{+0.09}_{-0.08}$ & $0.11^{+0.04}_{-0.04}$  & $>14.6,35.9$    & b       \\
 $187798$  & $0.38^{+0.4}_{-0.4}$     & $0.75^{+0.2}_{-0.3}$   & $8.94^{+0.4}_{-0.07}$  & $334^{+327}_{-156}$    & $0.17^{+0.2}_{-0.1}$   & $0.04^{+0.09}_{-0.03}$  & $>4.8,39.7$     & b       \\
 $191179$  & $0.06^{+0.02}_{-0.02}$   & $0.11^{+0.06}_{-0.04}$ & --                     & $48^{+14}_{-14}$       & --                     & --                      & $>0.5,0.6$      & sf      \\
 $193971$  & $0.09^{+0.05}_{-0.04}$   & $0.16^{+0.07}_{-0.06}$ & --                     & $7^{+2}_{-2}$          & --                     & --                      & $>0.5,0.6$      & sf      \\
 $213260$  & $0.75^{+0.2}_{-0.2}$     & $0.78^{+0.1}_{-0.2}$   & $8.96^{+0.1}_{-0.06}$  & $964^{+655}_{-408}$    & $0.13^{+0.2}_{-0.07}$  & $0.03^{+0.07}_{-0.02}$  & $>13.9,54.2$    & b       \\
 $229763$  & $0.15^{+0.08}_{-0.05}$   & $0.45^{+0.3}_{-0.2}$   & $8.61^{+0.3}_{-9}$     & $318^{+104}_{-143}$    & $1.09^{+0.3}_{-0.4}$   & $0.52^{+0.2}_{-0.2}$    & $>0.8,2.9$      & pb      \\
 $255581$  & $0.53^{+0.3}_{-0.2}$     & $0.62^{+0.2}_{-0.2}$   & --                     & $285^{+212}_{-126}$    & $0.22^{+0.3}_{-0.1}$   & $0.06^{+0.1}_{-0.04}$   & $>4.2,22.6$     & b \\\hline
 $65602$   & $0.61^{+0.3}_{-0.4}$     & $0.91^{+0.06}_{-0.1}$  & $8.93^{+0.02}_{-0.04}$ & $1213^{+1293}_{-616}$  & $0.20^{+0.3}_{-0.1}$   & $0.05^{+0.1}_{-0.04}$   & $>17.6,114.1$   & b       \\
 $96779$   & $0.18^{+0.3}_{-0.2}$     & $0.55^{+0.2}_{-0.2}$   & $9.09^{+0.1}_{-0.2}$   & $174^{+160}_{-93}$     & $0.16^{+0.3}_{-0.1}$   & $0.05^{+0.1}_{-0.03}$   & $>1.7,11.6$     & b       \\
 $98596$   & $0.21^{+0.4}_{-0.2}$     & $0.50^{+0.3}_{-0.3}$   & $9.12^{+0.2}_{-0.2}$   & $322^{+393}_{-199}$    & $0.15^{+0.3}_{-0.1}$   & $0.05^{+0.1}_{-0.03}$   & $>0.9,10.0$     & pb      \\
 $108587$  & $0.00^{+0.006}_{-2e-06}$ & $0.77^{+0.2}_{-0.2}$   & $8.93^{+0.2}_{-0.02}$  & $2418^{+1876}_{-1192}$ & $0.13^{+0.2}_{-0.08}$  & $0.03^{+0.09}_{-0.03}$  & $>9.1,45.7$     & b       \\
 $114152$  & $0.07^{+0.3}_{-0.07}$    & $0.57^{+0.3}_{-0.3}$   & $9.17^{+0.1}_{-0.3}$   & $651^{+754}_{-417}$    & $0.14^{+0.3}_{-0.09}$  & $0.04^{+0.1}_{-0.03}$   & $>0.7,13.8$     & pb      \\
 $116031$  & $0.40^{+0.4}_{-0.4}$     & $0.84^{+0.1}_{-0.2}$   & $8.83^{+0.09}_{-0.09}$ & $3437^{+1506}_{-1682}$ & $0.10^{+0.2}_{-0.05}$  & $0.02^{+0.06}_{-0.01}$  & $>13.0,78.0$    & b       \\
 $124662$  & $0.26^{+0.1}_{-0.09}$    & $0.57^{+0.3}_{-0.2}$   & $0.00^{+9}_{-0}$       & $29^{+9}_{-12}$        & $1.33^{+0.3}_{-0.3}$   & $0.44^{+0.2}_{-0.2}$    & $>0.7,2.6$      & pb      \\
 $125588$  & $0.31^{+0.2}_{-0.2}$     & $0.54^{+0.3}_{-0.2}$   & $8.92^{+0.2}_{-0.09}$  & $209^{+216}_{-106}$    & $0.16^{+0.3}_{-0.1}$   & $0.05^{+0.1}_{-0.03}$   & $>2.4,13.9$     & b       \\
 $136729$  & $0.20^{+0.2}_{-0.09}$    & $0.37^{+0.3}_{-0.2}$   & $8.74^{+0.1}_{-1}$     & $21^{+27}_{-11}$       & $0.38^{+0.4}_{-0.3}$   & $0.14^{+0.2}_{-0.1}$    & $>0.6,5.3$      & pb      \\
 $138600$  & $0.45^{+0.3}_{-0.2}$     & $0.58^{+0.2}_{-0.2}$   & $8.91^{+0.04}_{-0.1}$  & $773^{+863}_{-402}$    & $0.15^{+0.3}_{-0.1}$   & $0.05^{+0.09}_{-0.03}$  & $>0.9,14.6$     & pb      \\
 $162358$  & $0.23^{+0.2}_{-0.09}$    & $0.32^{+0.2}_{-0.1}$   & $8.77^{+0.1}_{-0.2}$   & $35^{+36}_{-16}$       & $0.22^{+0.3}_{-0.1}$   & $0.07^{+0.1}_{-0.05}$   & $>1.3,6.9$      & b       \\
 $165790$  & $0.05^{+0.03}_{-0.01}$   & $0.08^{+0.06}_{-0.03}$ & --                     & $35^{+13}_{-15}$       & --                     & --                      & $>0.5,0.5$      & sf      \\
 $212238$  & $0.29^{+0.5}_{-0.3}$     & $0.76^{+0.1}_{-0.2}$   & $8.92^{+0.2}_{-0.04}$  & $793^{+681}_{-379}$    & $0.16^{+0.3}_{-0.1}$   & $0.04^{+0.09}_{-0.03}$  & $>6.2,36.4$     & b       \\
 $240201$  & $0.29^{+0.5}_{-0.3}$     & $0.68^{+0.2}_{-0.5}$   & $8.94^{+0.3}_{-0.1}$   & $798^{+770}_{-403}$    & $0.16^{+0.2}_{-0.1}$   & $0.04^{+0.08}_{-0.03}$  & $>1.2,19.8$     & b       \\
\hline
\end{tabular}
    \label{tab:pipes2}
\end{table*}

\begin{figure*}
    \centering
    \includegraphics[width=\textwidth]{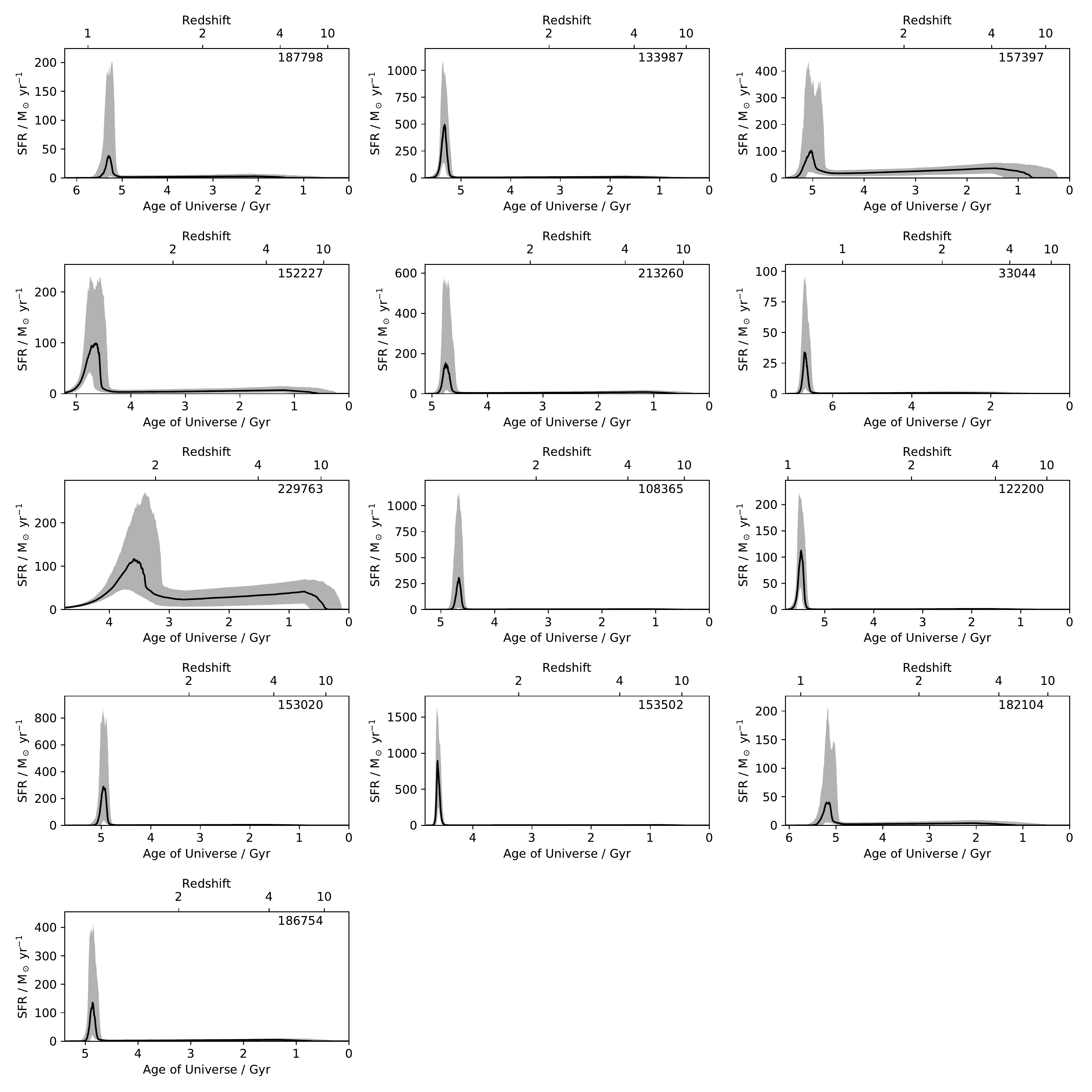}
    \caption{A montage of the fitted star formation histories for the galaxies classified as a post-starburst both spectroscopically and photometrically. The black line shows the median of the posterior SFH at each age, while the shading indicates the 16th and 84th percentile confidence intervals. The DR11 ID is given in the top right of each panel.}
    \label{fig:sfh_SCpsb}
\end{figure*}

\begin{figure*}
    \centering
    \includegraphics[width=\textwidth]{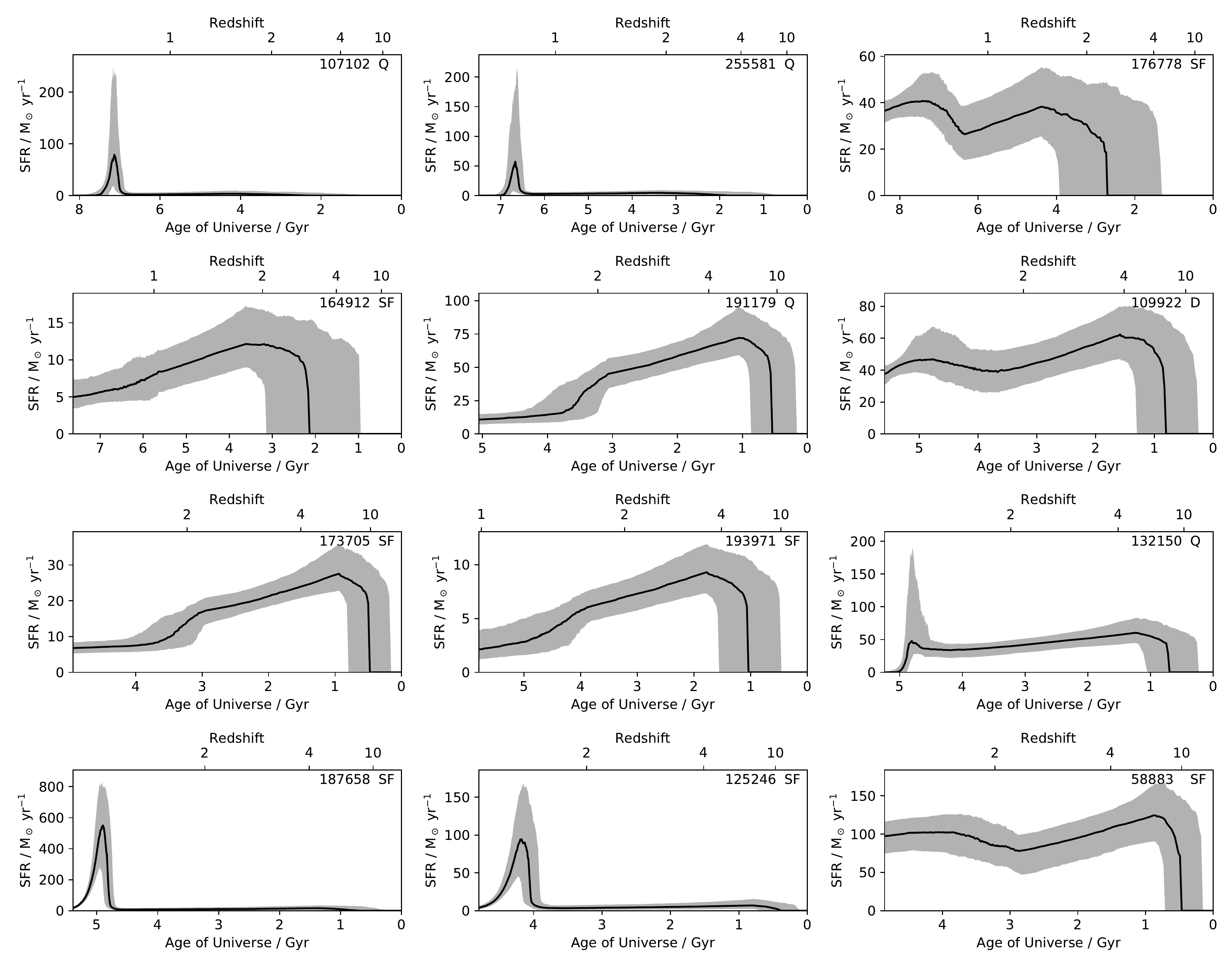}
    \caption{A montage of the fitted star formation histories for the galaxies classified as a post-starburst spectroscopically, but not photometrically. The lines are as in Fig.~\ref{fig:sfh_SCpsb}. The DR11 ID is given in the top right of each panel, as well as the super-colour class (Q = quiescent, SF = star forming, D = dusty). }
    \label{fig:sfh_psb}
\end{figure*}

\begin{figure*}
    \centering
    \includegraphics[width=\textwidth]{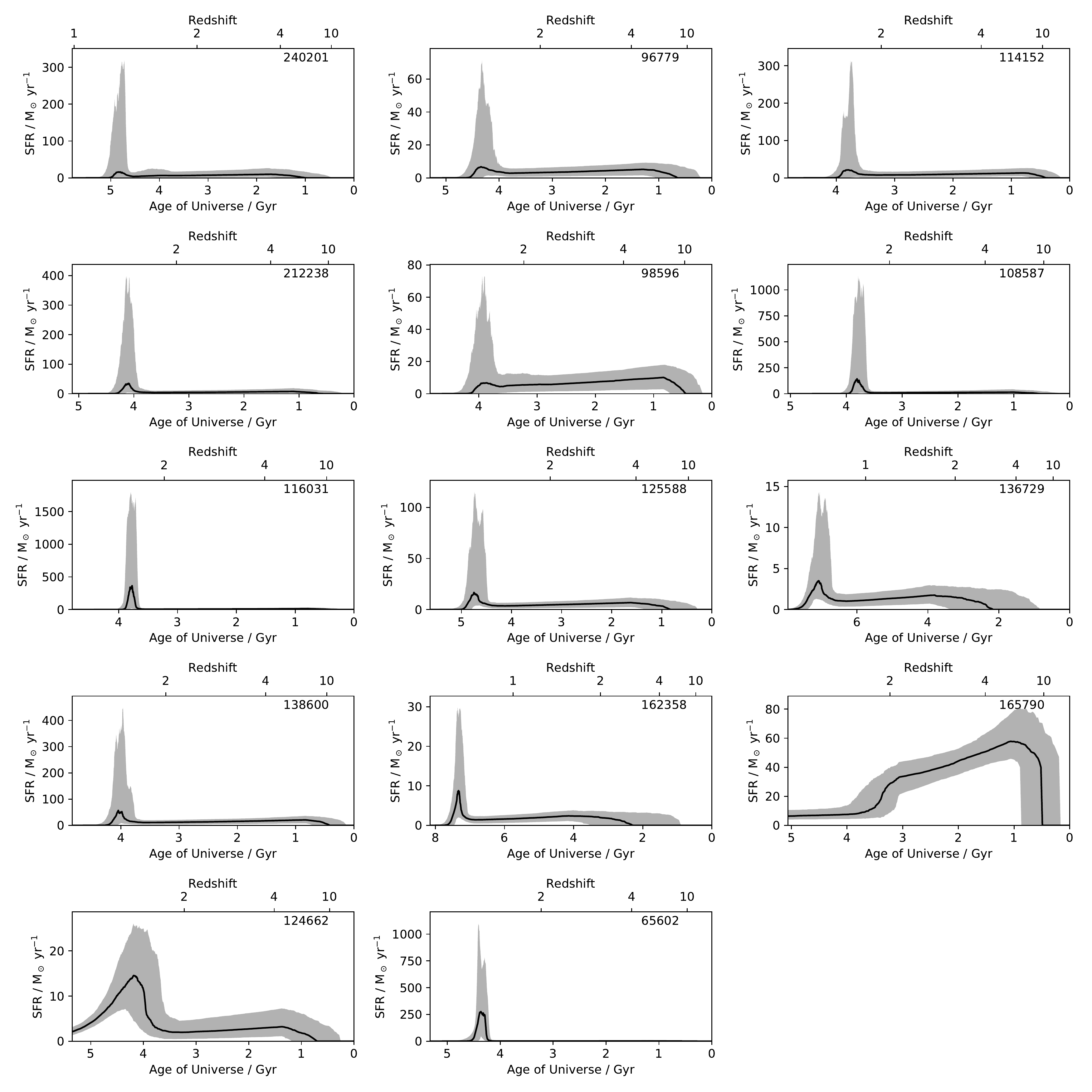}
    \caption{A montage of the fitted star formation histories for the galaxies classified as post-starbursts  photometrically but not spectroscopically. The lines are as in Fig.~\ref{fig:sfh_SCpsb}.}
    \label{fig:sfh_SCpsb_only}
\end{figure*}

\begin{figure*}
    \centering
    \includegraphics[width=0.75\textwidth]{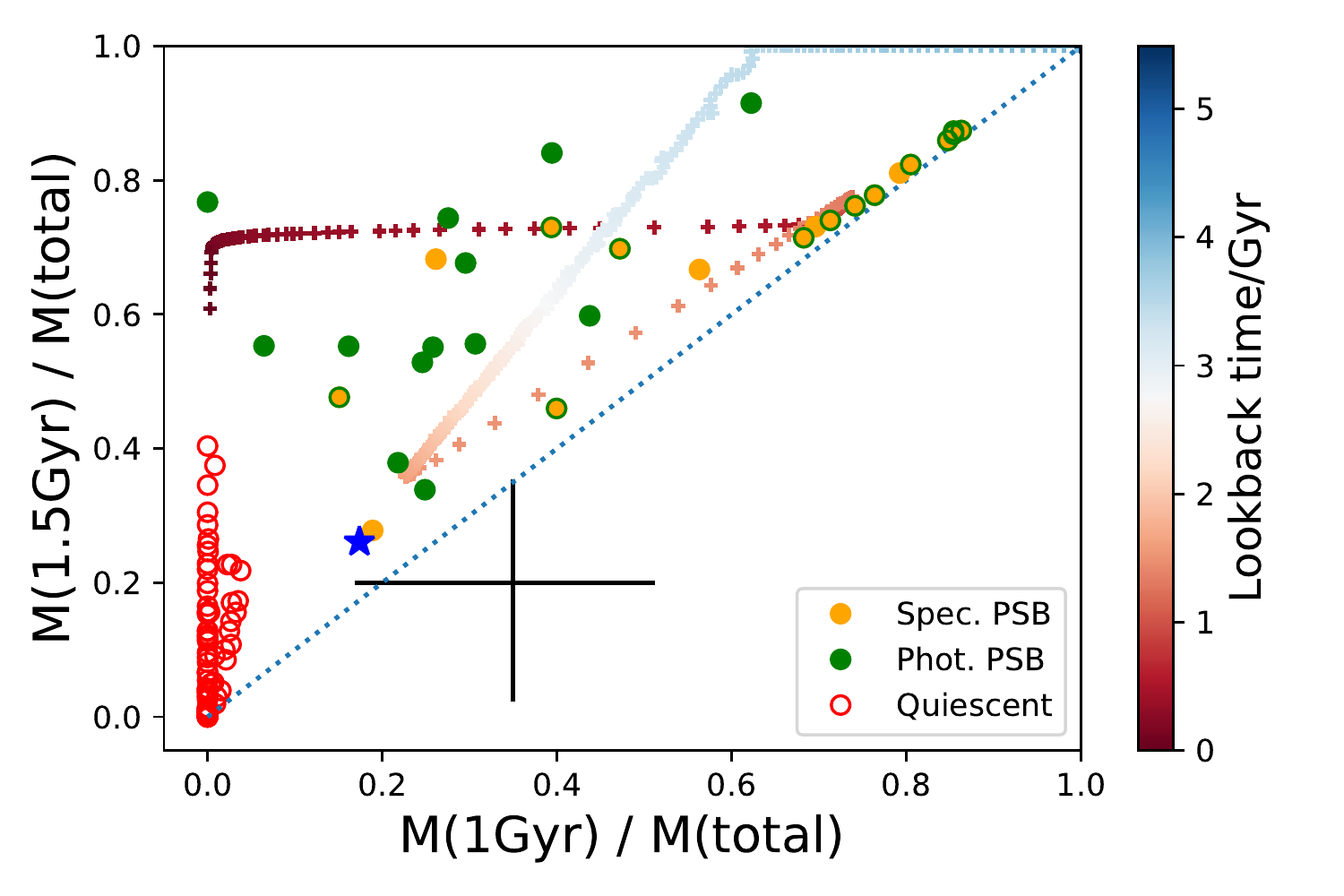}
    \caption{The fraction of stars formed in the last 1\,Gyr vs. 1.5\,Gyr for the post-starburst and quiescent samples. Only post-starburst galaxies with burst-like star formation histories are shown. Green (orange) points show post-starbursts identified photometrically (spectroscopically). Post-starbursts that are identified both photometrically and spectroscopically are marked as orange with green outer rings. The error bar shows the median error for the post-starburst galaxies. The small crosses, colour coded according to lookback time from $z=1$ shown in the right hand colour bar, show the evolutionary track taken by the SFH shown in Fig.\ref{fig:sfh} with points spaced equally in time every 10\,Myr, for a burst with an age of 1.5\,Gyr and burst mass fraction of 70\%. The blue star indicates the position of a star forming galaxy that has undergone constant star formation for 5.75\,Gyr (the age of the Universe at $z=1$).}
    \label{fig:f1f1p5}
\end{figure*}

\begin{figure}
    \centering
    \includegraphics[width=\columnwidth]{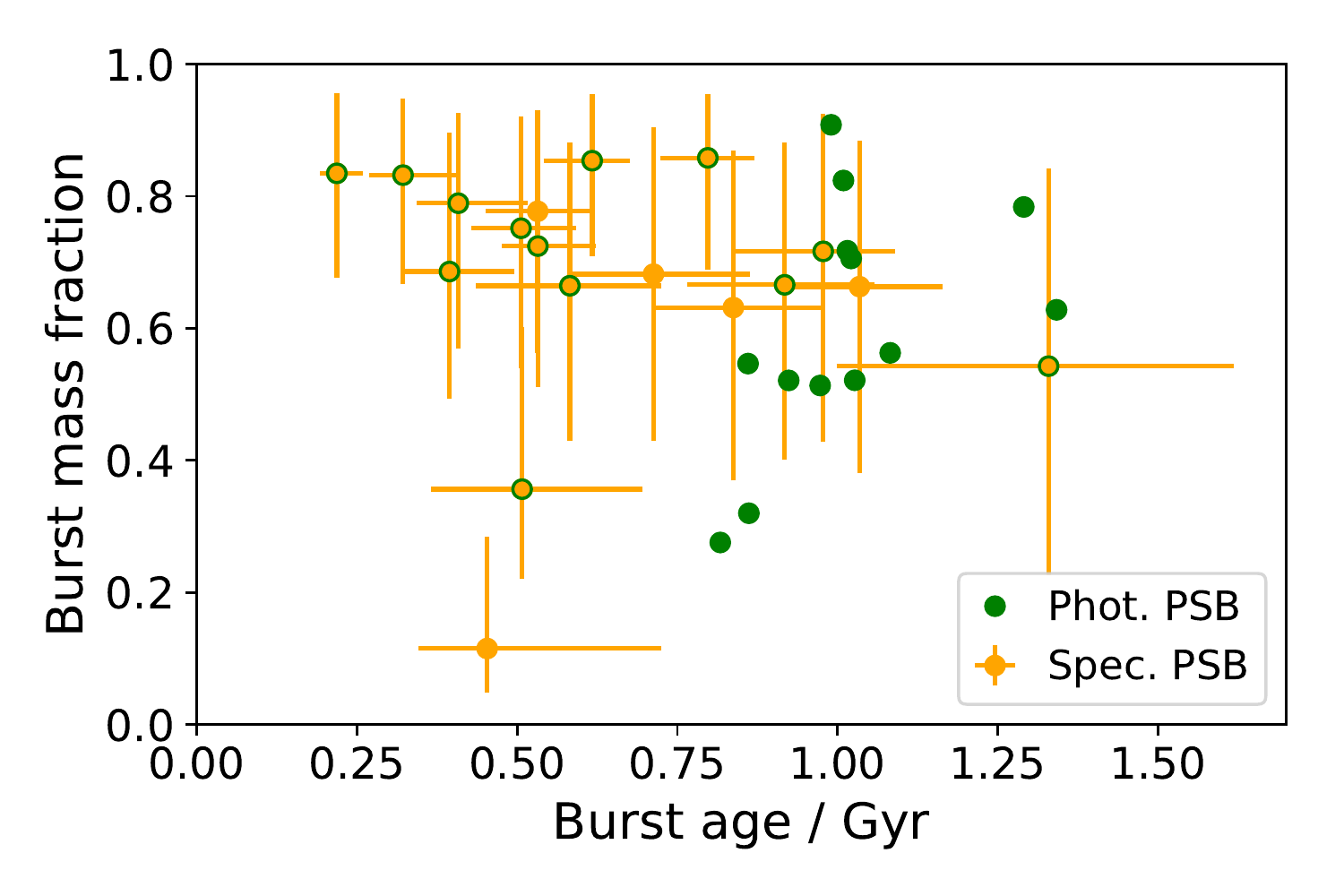}
    \caption{For those galaxies with burst-like star formation histories, we plot the burst mass fraction vs. burst age. Errors on the photometrically identified post-starbursts are omitted for clarity, but are similar to the other classes. Symbols are as in Fig.~\ref{fig:f1f1p5}.}
    \label{fig:fbursttburst}
\end{figure}

Fig.~\ref{fig:sfh_SCpsb} shows a montage of the posterior star formation histories for the galaxies identified as post-starburst galaxies both spectroscopically and photometrically (i.e. with strong Balmer breaks, weak nebular emission and strong Balmer absorption lines), Fig.~\ref{fig:sfh_psb} shows those identified purely spectroscopically, and Fig.~\ref{fig:sfh_SCpsb_only} shows those identified purely photometrically. Immediately we see a clear difference between the three samples: while the photometric criteria clearly identifies a majority of objects that most people would be happy calling ``post-starburst'' i.e. having experienced a recent short lived burst of star formation and rapid quenching, this is not true for the majority of the spectroscopic-only sample. We note that all the fits are formally good  (i.e. have a reduced chi-squared for the best-fit models of around unity), due in large part to the calibration and noise components included in the fits. To reassure ourselves that the SFHs are robust, we compared them to those derived from the double powerlaw fit, both with and without the Gaussian process noise, finding all galaxies to show qualitatively similar SFHs regardless of the priors.  In Tables \ref{tab:pipes1} and \ref{tab:pipes2}  we list the main properties of the galaxies from the {\sc bagpipes} fit, and relevant quantities derived from the fitted star formation histories. We find that the galaxies studied typically have stellar masses in the range $10^{10}-10^{11.25}M_\odot$, and stellar metallicities typically lie in the range $0.5-2$ times solar. We note that the spectra do not go red enough to contain the strong metal lines for accurate  metallicity constraints, so these values should be treated with caution. The remaining quantities will be described in detail in the following sections. 

In order to provide a quantitative way to distinguish ``burst-like'' SFHs from those with no evidence of a burst, we calculate the mean SFR during the burst (${\rm SFR}_{burst}$), $\pm50$\,Myr either side of $t_{burst}$, and compare this to the mean SFR between 200 and 300\,Myr prior to the burst (${\rm SFR}_0$) for 500 random draws from the posterior star formation histories. In the penultimate column of Table \ref{tab:pipes2} we present the 0.15th and 16th percentiles of the distributions of ${\rm SFR}_{burst}/{\rm SFR}_0$ as lower limits on this ratio. We classify as ``burst-like'' those where the 0.15th percentile is $>1$ and those as ``star-forming'' where the 16th percentile is $<1$. Those in between are classified as ``possible burst'' and these classifications are given in the final column of Table \ref{tab:pipes2}. We note that these quantitative classifications of the SFHs are very close to those that result from a simple visual inspection of the SFHs in Figs. \ref{fig:sfh_SCpsb}, \ref{fig:sfh_psb} and \ref{fig:sfh_SCpsb_only}. 

\subsection{Burst mass fractions and ages}

Fig.~\ref{fig:f1f1p5} shows the fraction of mass formed in the last 1\,Gyr vs. 1.5\,Gyr for the quiescent and post-starburst samples  with burst-like or possible burst-like star formation histories \footnote{Fraction of mass is defined as the integrated SFH within some time, divided by the total integrated SFH. It does not  account for mass loss.}, and these quantities are given in columns 2 and 3 of Table \ref{tab:pipes2}. There is a clear difference between the samples: very few of the quiescent galaxies have formed any stars in the last 1\,Gyr, and the majority having formed $<$20\% in the last 1.5\,Gyr; the post-starbursts classified photometrically have formed $>40$\% of their mass in the past 1.5\,Gyr and $<40$\% in the past 1\,Gyr; and the majority of post-starbursts classified both spectroscopically and photometrically have formed $>40$\% of their mass in the past 1\,Gyr. The blue star indicates the position of a hypothetical galaxy which has undergone constant star formation for 5.75\,Gyr, the age of the Universe at $z=1$. 

Overplotted on Fig.~\ref{fig:f1f1p5} is the evolutionary track taken by the SFH shown in Fig.~\ref{fig:sfh}, for a burst mass fraction of 70\% and a burst age of 1.5\,Gyr. The track is colour coded by lookback time to the Big Bang for a $z=1$ galaxy and points are spaced equally in time every 10\,Myr. For an exponentially declining or constant SFH occurring before the starburst, galaxies evolve from top right first horizontally, then diagonally towards the bottom left on the diagram, as mass is steadily built. As the burst occurs, galaxies evolve up and right, meeting the group of spectroscopically and photometrically selected post-starbursts positioned in the upper right of the diagram. The evolution slows for a while, explaining the build up of observed galaxies forming a tight line here. As the burst ages, galaxies evolve horizontally, first rapidly then slowly, passing through the photometrically selected post-starbursts at a height determined directly by the burst mass fraction. Again the evolution slows, while there are very few observed galaxies here. This may be due to incompleteness in our sample, as bursts this old are hard to identify as post-starburst. Alternatively it may indicate that our model is too simple to describe the entire dataset. For clarity in the figure we stop the evolution at 1.5\,Gyr; from this point they evolve rapidly vertically downwards, reaching the origin 1.55\,Gyr after the burst. 

Fig.~\ref{fig:fbursttburst} shows the fraction of mass formed in the starburst component of the fitted SFH ($f_{burst}$) vs. the age of the starburst ($t_{burst}$, columns 3 and 4 in Table \ref{tab:pipes1}). As the SFH model is flexible enough to use the ``burst'' component to extend the exponential declining component, rather than creating a well defined ``burst'' of star formation, we restrict this to only those galaxies with burst-like or possible burst-like star formation histories based on the peak to pre-burst SFR ratio in Table \ref{tab:pipes2}, as described above. We measure total burst mass fractions of typically 40-90\%, and the onset of the burst occurs within the last 0.25-1\,Gyr for those identified spectroscopically, and slightly earlier (0.8-1.3\,Gyr) for those identified only with photometry. This older burst in the sample selected purely photometrically is consistent with their weaker \hda\ absorption lines which causes them to be missed from the spectroscopic selection (see Fig.~\ref{fig:sample}).

\subsection{Further useful quantities}
Table \ref{tab:pipes2} provides the historical maximum star formation rate in the last 2\,Gyr and the quenching timescales calculated from 500 draws from the posterior star formation history distribution. The quenching timescales are provided in two different ways: firstly the time from the peak of the SFR to where the sSFR falls below $0.2/t_H$, where $t_H$ is the age of the Universe at the redshift of the quenching; secondly from where the sSFR falls from $1/t_H$ to $0.2/t_H$. The former measurement is designed to show how rapid ``rapid'' quenching needs to be, to match the observations of Balmer absorption lines and break in post-starburst galaxies. We see that values of 100--200\,Myr are typical for our samples. The latter measurement will allow us to compare directly with recent simulation results in the discussion section below. The $0.2/t_H$ criteria is a standard cut-off for identifying quenched galaxies used in the literature \citep[e.g.][]{Pacifici2016}. 

\subsection{Interlopers and escapees}

\begin{figure}
    \centering
    \includegraphics[width=\columnwidth]{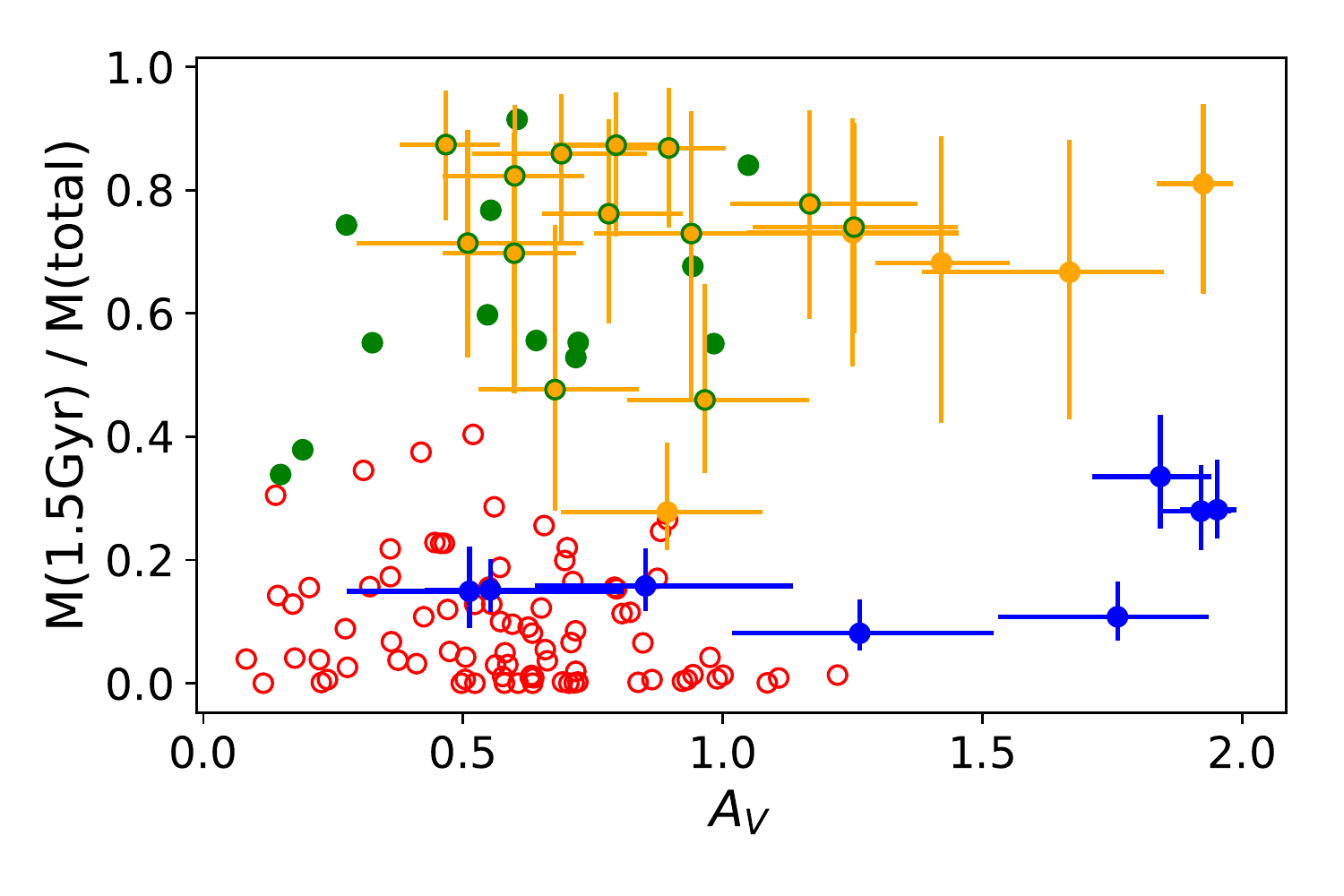}
    \caption{The ISM dust content ($A_V$) vs. the fraction of mass formed in the last 1.5\,Gyr. Symbols are as in Fig~\ref{fig:f1f1p5}, with the addition of blue filled circles which indicate the galaxies selected as post-starburst, but with SFHs that indicate they are in fact star-forming.}
    \label{fig:dust}
\end{figure}

We now turn to look at the objects which do not fit nicely into the picture. Firstly, there are either 4 or 5 post-starbursts with burst-like SFHs that are not classified as post-starbursts by the super-colours (DR11 IDs 107102, 255581, 187658 and 125246, 132150). This $\sim$15\% incompleteness is a little lower than reported in \citet{Maltby2016}. Secondly, we find that the spectroscopic post-starburst sample contains 7 objects with no apparent recent starburst (DR11 IDs 176778, 164912, 191179, 109922, 173705, 193971, 58883). 

We begin with the missing post-starbursts from the photometric selected sample. 132150 has a much lower burst mass fraction, which might explain why it is not identified by a photometric method, as we expect photometry to be less sensitive to weaker bursts than spectroscopy. 187658 and 125246 have slightly longer quenching times and are the two objects with significantly higher W[OII] than the rest of the sample, which might indicate their burst has not yet completely quenched\footnote{Additionally, 125246 is the only x-ray source in our PSB sample, so the [OII] may also arise from an AGN}. Both factors may cause a reduction in the strength of Balmer break needed for photometric selection. The penultimate column of Table \ref{tab:pipes1} provides the median and percentile fitted values for attenuation by dust.  Note that in our 2-component dust model, $A_V$ refers to the amount of dust affecting stars older than $10^7$ years. Younger stars are attenuated by an additional factor. We see that 107102, 255581 and 187658 have an unusually high fitted $A_V$ of 1.42, 1.66 and 1.93 magnitudes respectively, and 125246 is equally on the high side with $A_V=1.27$. These 4 objects with high dust contents where the photometry fails to identify the post-starburst nature of the galaxy fall in or very close to the super-colour defined red-sequence. Dust causes the super-colours to be biased to redder overall colours, and the distinctive triangular shape of the AF-star SED is lost.

In Fig.~\ref{fig:dust} we show the fitted attenuation by dust of the post-starburst and quiescent samples against the fraction of mass formed in the last 1.5\,Gyr. The majority of quiescent galaxies have $A_V$ in the range $0-1$\,mag. The post-starburst galaxies classified both spectroscopically and photometrically are fitted with a slightly higher $A_V$ in the range $0.5-1.2$\,mag as might be expected given the more recent star-formation activity\footnote{We note that for the post-starburst sample alone there is only a very weak correlation between burst age and $A_V$ with a Pearson correlation coeffecient of -0.1.}. However, as well as the 4/5 of the post-starbursts not identified by the photometry, 4/7 of the spectroscopic-only post-starbursts with non burst-like SFHs also have $A_V>1.2$. Both results make intuitive sense: relying on W[OII] to identify galaxies with unusually low SFRs will also identify star-forming galaxies with unusually high dust contents; and even high quality multi-wavelength photometry can suffer from the well-known age-dust degeneracy. 

This leaves only 3 unexplained objects with normal dust contents where the strong Balmer absorption lines and weak [OII] suggest there has been a recent starburst, but the photometry and spectral fitting disagree. Inspecting the fits shows that there is excess UV flux in these objects compared to the other PSBs (164912, 173705 and 193971). Another factor that might cause the false identification of post-starburst galaxies in both photometry and spectroscopy is the relative geometry of dust and stars which we attempted to fit for with the $\eta$ parameter which controls the additional fraction of dust attenuating young stars. OB stars can be hidden behind dense dust clouds, rather than absent from the galaxy, causing stronger Balmer lines than expected. Unfortunately this quantity is not well constrained with the current data and in most cases the posterior distribution is close to the prior. However, 2 of these 3 do show some evidence for enhanced $\eta$. Only improved data will allow us to ascertain whether these are truly post-starburst galaxies or interlopers caused by dust. Finally, there is always the possibility of unseen problems with the data. For example, object 193971 has a lower SNR spectrum than typical for the sample, as well as a close neighbour which may affect the accuracy of the photometry.

\subsection{Parametric vs. non-parametric star formation histories}

There are known problems with parametric SFHs, in particular systematic underestimation of mass weighted age, and tight prior distribution on sSFR \citep{CarnallLeja2019}.  A second common approach to reveal the star formation histories of galaxies is through ``non-parametric'' fitting where simple stellar populations (SSPs) are added in linear combination. This is typically used where data quality is higher, but a comparison between the two approaches is valuable. One popular example of a non-parametric code is {\sc starlight} \citep{Starlight2005}, which has recently been adapted to simultaneously fit photometry and spectroscopy \citep{LopezFernandez2016,Werle2019}. However, as {\sc starlight} has predominantly been used by the low-redshift community on high quality SDSS spectra, there are no allowances for relative spectrophotometric calibration errors or systematic noise components as in {\sc bagpipes}.

In Appendix \ref{App:starlight} we present a comparison between {\sc starlight} fits and {\sc bagpipes} fits to two of our post-starburst galaxies. Despite the large difference between the codes, we see a reasonable agreement in terms of fraction of mass formed in the last 1 and 1.5\,Gyr. Overall, we find that {\sc bagpipes} produces a slightly smaller range of burst mass fractions compared to {\sc starlight}. However, without allowing for the additional calibration errors and systematic noise that are known to exist in these spectra, a robust quantitative comparison is unfortunately not possible. 

\section{Discussion}\label{sec:discussion}
In this section we discuss our results in relation to the importance of rapid quenching for building the red sequence, possible progenitors and implication of the quenching timescales for feedback mechanisms. Finally, we address the question of the value added by ``expensive'' spectroscopy in this era of massive high quality multi-wavelength datasets. 

\subsection{``Post-starburst'' or ``rapidly quenched''}
Whether or not a significant starburst precedes the rapid quenching of star formation has important implications for the cause of the quenching event. The gas inflows caused by a gas rich merger are expected to lead to a significant burst of star formation, that may subsequently quench by \emph{internal} processes, such as gas exhaustion or expulsion by SNe driven winds or AGN. On the other hand, a galaxy falling into a cluster environment may be rapidly stripped of its gas by \emph{external} ram pressure without a (or with a lesser) preceding starburst. Other processes are less clear cut, with harassment by neighbouring galaxies potentially also leading to gas inflows and a starburst, although this may be weaker than in the case of nuclear coalescence. 

The penultimate column of Table \ref{tab:pipes2} gives the 0.15th and 16th percentiles\footnote{Corresponding to $3\sigma$ and $1\sigma$ lower limits in the case of a Gaussian distribution.} of the ratio of SFR during the burst, compared to before the burst. Where this is $>1$ there is evidence of a detectable rise in star formation prior to quenching. We find 24 (31) have a detectable rise in their SFR (i.e. ${\rm SFR}_{ratio}>1$) that is significant at $\gtrsim 3\sigma$ ($1\sigma$). Additionally, 26 have SFR rises of $>10$ at $\gtrsim 1\sigma$.  This certainly implies that a significant fraction of our galaxies are truly ``post-starburst'' rather than simply ``rapidly quenched'', particularly those identified photometrically. On the other hand, 3 galaxies (191179, 173705, 165790) do show some signs of a truncated star formation history, with 2 of these not identified photometrically. We caution that the stronger the burst, the stronger the spectral features and our results do not exclude  the existence of ``rapidly quenched'' galaxies that we are either unable to select or unable to identify the quenching in their SFH with the current data available at $z\sim1$.

The measured burst mass fractions of many of these post-starburst galaxies are arguably more consistent with being the dominant ``formation'' episode of the galaxy, rather than a dramatic event occurring during their otherwise normal evolution. The burst mass fractions are far larger than the  ``rejuvenation'' events detected in the spectroscopic Lega-C survey \citep{Chauke2019}, but consistent with the ``Late bloomers'' identified in the prism observations of the Carnegie-Spitzer-IMACS study \citep{Dressler2018}. Whether such dramatic formation events at $z\sim1$ are consistent with the current generation of hydrodynamic galaxy evolution simulations is clearly a matter that demands urgent investigation. 

\subsection{Time spent in the post-starburst phase, and role of post-starbursts in red sequence growth}

In \citet{Wild2009} and \citet{Wild2016} we summed the total amount of mass in galaxies caught during the post-starburst phase, and compared this to the amount of mass growth of the red sequence at a similar or slightly later epoch. The results suggested that a significant fraction of new galaxies entering the red sequence might have recently undergone a period of fast growth and rapid quenching. If these starburst events are caused by gas rich major mergers, this is at odds with similar estimates from close pair and major merger fractions \citep[e.g.][]{LopezSanjuan2010,Weigel2017}. The primary uncertainty in this calculation is the unknown ``visibility'' time for the post-starbursts i.e. how long a galaxy would be observed as a post-starburst for. Comparison of the amplitudes and shapes of mass functions of quiescent and post-starburst galaxies suggested that up to 100\% of the red sequence galaxies at $z\sim0.5-1.5$ could have passed through a post-starburst phase, which would imply that a rapid shut down in star formation is the prevalent form of quenching at these redshifts. However, for this to hold, very short visibility times of $\sim$250\,Myr were required. \citet{Belli2019} revisited this calculation with spectroscopic information allowing star formation histories of 24 quiescent galaxies to be constrained, a few of which lie in the post-starburst region of the UVJ diagram. They used the median ages of the stellar populations to conclude that fast quenching accounts for only about a fifth of the growth of the red sequence at $z\sim1.4$. 

Taking our sample of super-colour selected post-starburst galaxies with spectra, we use 500 draws from the posterior star formation histories to calculate 500 evolutionary tracks through super-colour space. From these we calculate the time between them entering and leaving the post-starburst region of the diagram. 
Table \ref{tab:pipes2} gives the posterior median visibility times for galaxies identified as post-starburst photometrically (i.e. those relevant to the calculation of \citet{Wild2016}). We see that the majority of the star formation histories indicate visibility times of 500\,Myr -- 1Gyr, with no difference depending on whether they are selected photometrically or not. This is similar to the value of $\sim600$\,Myr estimated from isolated galaxy merger simulations by \citet{Wild2009}. Combining with the results of \citet{Wild2016} we conclude that 25\%-50\% of the growth of the red sequence at $z\sim1$ is caused by rapid quenching with galaxies passing through a post-starburst phase. This remains a little higher than \citet{Belli2019}, but probably consistent given the errors inherent in the extraction of star formation histories, and small number statistics in their sample. From our combined spectroscopic and photometric sample we can additionally revisit the incompleteness of the super-colour selected sample, finding $\sim15$\% (4-5/31) of true post-starburst galaxies are missed by this selection method. This would increase the fraction of red sequence growth through a post-starburst phase by a small amount. We emphasise that these results only strictly apply to red-sequence growth at redshifts slightly below one. It is plausible that the visibility times vary systematically with redshift if the physical processes leading to the starburst and quenching change. High quality spectroscopic data of large samples of objects at $z>1$ are required to reveal whether this is the case. Naturally, these red sequence growth rate fractions all hinge upon the fact that the post-starburst galaxies will not be rejuvenated in the near future, which is obviously impossible to assess with this data alone. Rejuvenation events are much more difficult to detect directly than quenching events, due to the brightness of the highest mass stars obscuring the older stellar populations \citep{Chauke2019}, and comparison with models may be the most promising avenue for progress in this direction \citep{Pandya2017,Behroozi2019}.

We briefly note that the fitted star formation histories of many of the galaxies lead to some evolutionary colour tracks that have significantly higher maximum SC2 values (larger Balmer breaks) than observed in our full photometric samples \citep{Wild2016}. These have the longest visibility times, and therefore we may be slightly underestimating the contribution of post-stabursts to the growth of the quiescent population. A combined analysis of the entire sample of galaxies may be able to improve on this result in the future.

\subsection{Possible progenitors}

In the local Universe, post-starburst galaxies are often presented as the natural descendants of local ultra-luminous infrared galaxies (ULIRGs), which are typically caused by major gas-rich mergers. Two possible progenitors of high redshift post-starburst galaxies are galaxies detected in the sub-mm or compact star forming galaxies (cSFGs).

In \citet{Wild2016} we noted that the post-starburst galaxies have a similar characteristic stellar mass, and a space density $6-7$ times higher than sub-mm galaxies \citep{Swinbank:2014,Simpson:2014}, which would lead to a direct correspondence if the visibility time was $6-7$ times shorter for sub-mm galaxies. For visibility times of 0.5-1\,Gyr measured from the fitted star formation histories, this would imply visibility times of 80-150\,Myr for galaxies in a sub-mm phase which are consistent with those inferred by \citet{Hainline2011} and \citet{Hickox2012}. In Table \ref{tab:pipes2} we present the maximum historical SFR of the fitted star formation histories for each post-starburst galaxy in the last 2\,Gyr, finding that all the galaxies with burst-like SFHs have median values of several hundred $M_\odot$/yr, while a small tail extends out to several thousand $M_\odot$/yr. Assuming a standard conversion from $870\mu m$ flux and total far-infrared luminosity, \citet{Swinbank:2014} find a range of SFRs of 20-1030$M_\odot$/yr with a median of 310$M_\odot$/yr for sub-mm galaxies. An upper cut-off in star formation rate is expected at $\sim1000 M_\odot$/yr, once duplicity of sources is accounted for \citep{Karim2013,Simpson2015}. This very close correspondence between the inferred starburst strengths of the post-starburst and sub-mm galaxies is perhaps surprising, given the errors inherent in both analyses, however once again affirms the likely close connection between these two populations. 

Compact star forming galaxies have also been suggested as possible progenitors of post-starbursts. These galaxies are detected in optical and NIR surveys with high stellar surface mass densities and significant amounts of obscured star formation \citep{Barro2013,Barro:2014,vanDokkum:2015p10009}.  \citet{Barro2013} find SFRs of $100-200M_\odot/yr$ and number densities of $\sim8\times10^{-5}$ at $1.5<z<3$ for cSFGs with $M*>10^{10}M_\odot$. For post-starbursts with $1<z<2$ and $M*>10^{10}M_\odot$ \citet{Wild2016} measure number densities of $\sim5-6\times10^{-5}$\,Mpc$^{-3}$.  Therefore, for all cSFGs to become post-starbursts would require the starbursts to have similar visibility times to the post-starburst phase, i.e. 500\,Myr -- 1Gyr, which is inconsistent with our fitted star formation histories and also with their inferred gas depletion times of $\sim100$\,Myr \citep{Barro2016}. If cSFGs are shorter lived, but with a similar number density, then only a fraction of them can be progenitors of post-starbursts. We note that  $z>1$ post-starbursts are also significantly more compact than quiescent galaxies at the same redshift \citep{Almaini2017}, compared to cSFGs which have similar or slightly larger sizes to quiescent galaxies of the same stellar mass. Thus it is possible that post-starbursts are descendants of only the most extreme processes that lead to cSFGs. 

\subsection{Quenching timescales and AGN feedback}
The timescale over which a galaxy quenches is thought to provide strong constraints on the physical process that causes the quenching, with feedback from AGN in particular expected to lead to the most rapid quenching events. Post-starburst galaxies are particularly relevant for identifying possible occurances of AGN feedback, due to the fast quenching times needed to cause the strong Balmer absorption and Balmer break features. 

We find that the time from the peak of the starburst to the point where the sSFR falls below $0.2/t_H$, where $t_H$ is the age of the Universe at the time of the quenching, is typically 100-200\,Myr (Table \ref{tab:pipes2}). In order to compare directly to simulations, we also calculated the time taken for the galaxy's sSFR to fall from $1/t_H$ to $0.2/t_H$, following \citet{Curro2019} for the {\sc simba} cosmological hydrodynamic simulations \citep{Simba2019}. They noted strongly bimodal quenching times, split at $\tau_q/t_H = 0.03$, with the fast quenching events ($\tau_q/t_H \sim 0.01$) likely caused by {\sc simba}'s jet-mode black hole feedback. We find that the majority of the post-starbursts have quenching times of 30-50\,Myr ($\tau_{q2}$ in Table \ref{tab:pipes2}), which corresponds to $\tau_q/t_H \sim 0.01$ at cosmic times of 4-5\,Gyr where most of our quenching events take place, exactly as found for the fast quenching mode in {\sc simba}. Interestingly, \citet{Curro2019} note that neither fast nor slow quenching events are directly linked to galaxy major mergers, where mergers are identified by a sudden increase in the stellar mass of the galaxy by more than 20\% (mass ratio $>1:4$). Interestingly, this is in contrast to \citet{Pawlik2019} and \citet{Davis2019} who found that \emph{local} ($z\sim0$) post-starburst galaxies in the {\sc eagle} cosmological hydrodynamic simulation were predominantly caused by major mergers or multiple minor mergers. A careful side-by-side comparison of post-starburst galaxies in observations and simulations, selected using the same observational techniques, is clearly warranted to uncover their possible causes at different redshifts, environments and stellar masses. 

We caution that while AGN feedback is currently the only possible mechanism for creating such rapid galaxy-wide shut-offs in star formation in simulations, the resolution of the simulations is not yet sufficient to resolve the star formation and interstellar medium, and at intermediate masses at least this may be sufficient to reduce or remove the need for efficient expulsion of gas by an AGN  \citep{NaabOstriker2017}. In this small spectroscopic sample only one galaxy has an X-ray detection indicating the presence of an AGN; larger spectroscopic samples would allow us to test for trends in quenching times with AGN signatures. NIR spectra would also allow us to identify obscured AGN through their rest-frame optical emission line ratios. The AGN content of super-colour selected post-starburst galaxies will be explored further in Almaini et al. (in prep). 

\subsection{Physical parameters with and without spectroscopy}
\begin{figure*}
    \centering
    \includegraphics[width=\columnwidth]{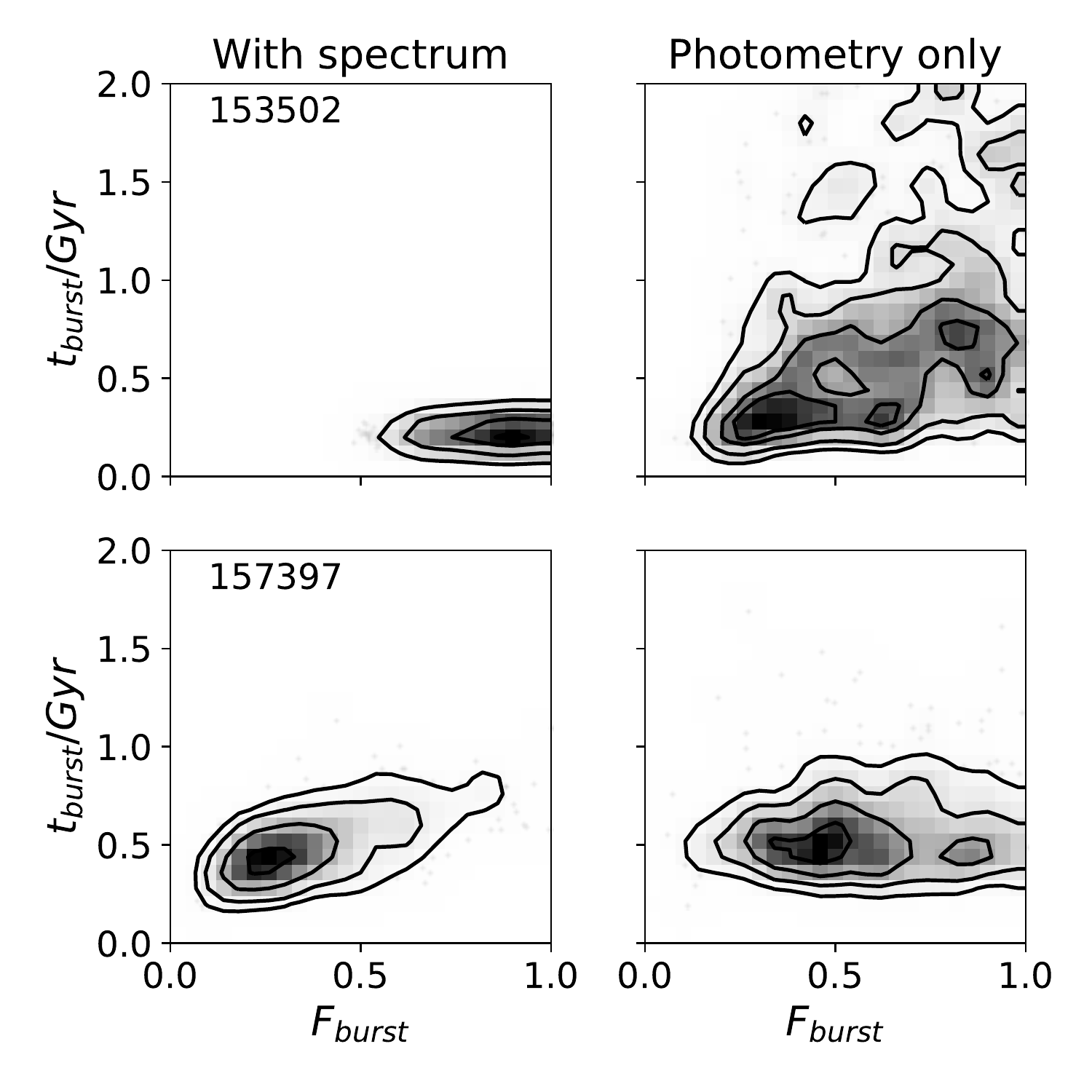}
    \includegraphics[width=\columnwidth]{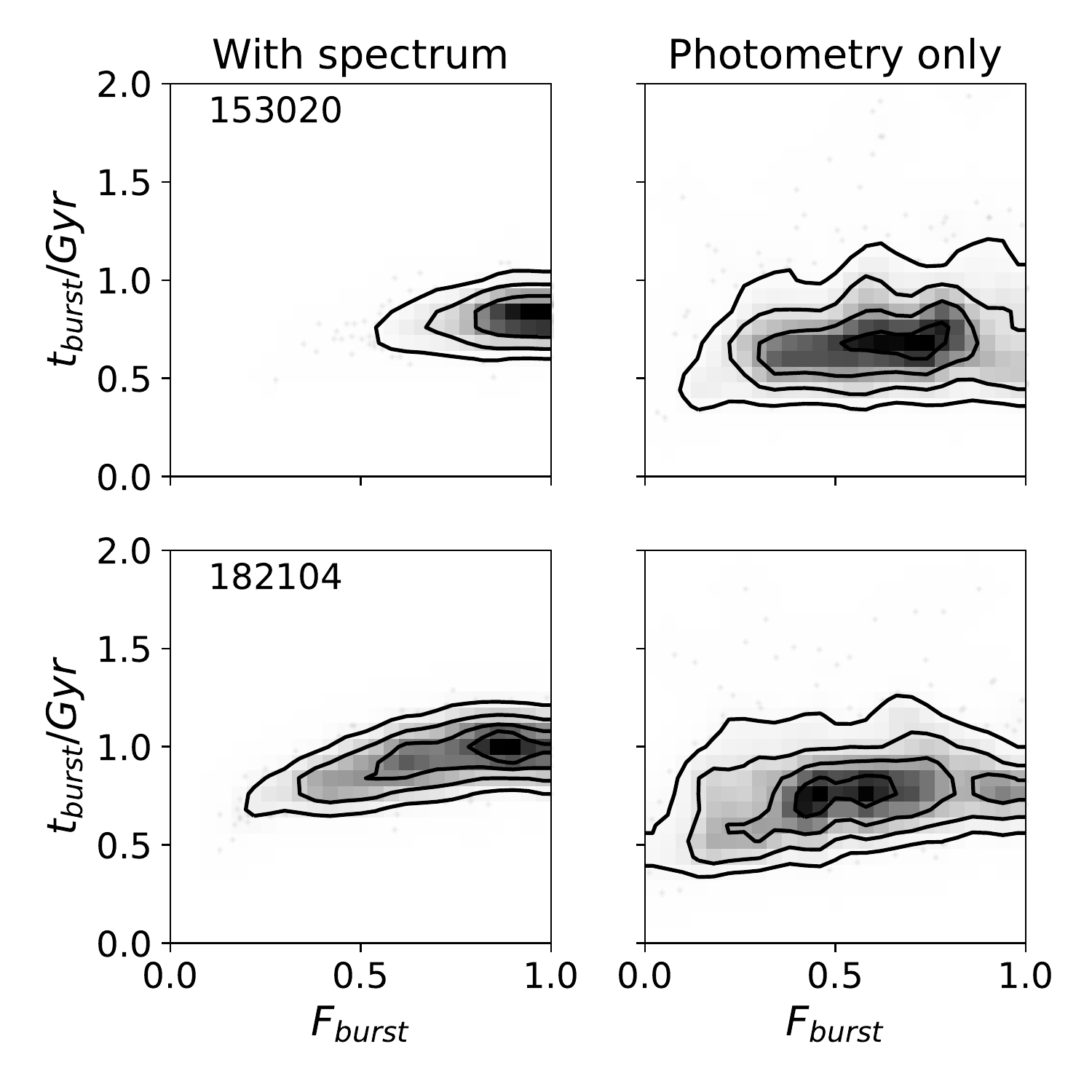}
    \caption{Comparison of the joint posterior distributions for the burst mass fraction and burst age, with and without the addition of spectroscopy, for 4 selected post-starbursts with a range of burst ages. We see that with only photometric information, the burst mass fraction is biased towards the median of the prior, and the burst ages are correspondingly biased due to the burst age-mass degeneracy.}
    \label{fig:specphot}
\end{figure*}

It is worthwhile questioning the need for follow-up spectroscopy in this era of very high quality, extensive, multi-wavelength photometric datasets. We run the same model fit again for all our post-starburst galaxies, but this time without including spectroscopy. In Fig.~\ref{fig:specphot} we compare the posterior distribution of the burst mass fraction and ages obtained for 4 post-starburst galaxies with a range of fitted burst ages, with and without spectroscopy included in the fit. Firstly, we notice how the parameters are less well constrained overall if spectroscopy is not available. Secondly, the burst mass fractions are biased towards the median of the prior (which is flat, with values between 0 and 1). Old, strong bursts have similar SEDs to young, weaker bursts, therefore the burst age is correspondingly biased. This highlights the importance of high SNR continuum spectroscopy for constraining the physical properties of galaxies, even where multiwavelength photometry provides an efficient identification method.  

\section{Summary}

We identified 39 galaxies with $0.5<z<1.3$, $M*\gtrsim10^{10}M_\odot$ and strong Balmer break and/or Balmer absorption lines, either using broad band photometric ``super-colour'' selection or traditional spectroscopic W[OII]-\hda\ selection. By fitting both high quality spectroscopy and photometry using the {\sc bagpipes} spectral fitting code, we derived star formation histories for the galaxies, comparing to a control sample of quiescent galaxies. A summary of our results is as follows:
\begin{itemize}
    \item High quality photometric data can identify a wide range of post-starburst galaxies, without the need for spectroscopic data. We gain only 4 or 5/31 new post-starburst galaxies when we consider spectroscopic information, giving a photometric sample incompleteness of $\sim15\%$. These are misidentified by the photometry either due to low burst mass fractions (1/5), ongoing star formation (2/5) or high dust contents (4/5). 
    \item We show that high quality spectroscopic continuum data is required to accurately constrain the burst masses and ages of the post-starburst galaxies. Fitting of photometry alone leads to biased results that are strongly dependent on the assumed priors, largely due to the burst mass-age degeneracy. 
    \item We find that the photometric identification is better at identifying galaxies with older bursts than selection by \hda\ and W[OII]. This is not expected to be true at lower redshift where errors on spectral measurements are smaller, and H$\alpha$ is typically available, making it significantly easier to detect older or weaker bursts, truncation and rejuvenation events. 
    \item Using \hda\ and W[OII] alone to identify post-starburst galaxies at these redshifts leads to a significant contamination rate, with {\sc bagpipes} finding no evidence for a starburst in 7/25 objects. 4 of these possible interlopers have a high dust content, which presumably causes a small W[OII]. Two have some evidence that a high fraction of dust obscures the young stars, which would lead to stronger observed Balmer absorption lines, although better data is needed to confirm this. Contamination of the photometric sample with objects showing no evidence for a recent starburst is small (1/27). 
    \item The post-starburst galaxies have burst mass fractions of 40-90\%, and ages $\lesssim1.3$\,Gyr. When plotting the fraction of mass formed in the last 1 and 1.5\,Gyr, the distribution of observed post-starburst galaxies is well matched by the expected evolution of a galaxy that has undergone a recent burst of star formation.
    \item We find that the visibility times for the super-colour measured post-starburst features are 0.5-1\,Gyr, leading to post-starburst galaxies potentially accounting for 25-50\% of red sequence growth at these redshifts. 
    \item The maximum historical star formation rate estimated from the fitted star formation histories is consistent with post-starbursts being descendants of sub-mm galaxies. Their number densities and visibility times imply that the sub-mm phase is visible for 80-150\,Myr. On the other hand, the number density and visibility time of the post-starbursts rules out a 1:1 connection with compact SFGs detected in the optical and NIR, with post-starbursts being too rare to account for all cSFG galaxies.
    \item The measured quenching timescales are identical to the ``fast'' quenching mode identified in the {\sc simba} cosmological hydrodynamic simulation, which is expected to be caused by the jet-mode AGN feedback. 
\end{itemize}

There is a clear need for high quality continuum spectroscopy of many more $z>1$ galaxies in order to elucidate the range of mechanisms responsible for the quenching of galaxies at different redshifts and in different environments. Upcoming high multiplex instruments such as the Multi-Object Optical and Near-infrared Spectrograph \citep[MOONS,][]{MOONS2014} on the VLT and Prime Focus Spectrograph \citep[PFS,][]{PFS2014} on the Subaru telescope should provide the large numbers of sufficiently high quality spectra to continue and improve upon studies such as this one. 

\section*{Acknowledgements}

The authors acknowledge helpful discussions, advice and comments from Gustavo Bruzual, Stephane Charlot, Natalia vale Asari and Nimish Hathi. LTA acknowledges support from the Ministry Of Higher Education and Scientific Research (MOHESR), Iraq. AW acknowledges financial support from the Royal Society Newton Fund (grant NAF/R1/180403, PI Natalia Vale Asari) and Funda\c{c}\~{a}o de Amparo \`{a} Pesquisa do Estado de S\~{a}o Paulo (FAPESP) process number 2019/01768-6. We extend our gratitude to the staff at UKIRT for their tireless efforts in ensuring the success of the UDS project. This work is based in part on observations from ESO telescopes at the Paranal Observatory (programmes 094.A-0410, 180.A-0776 and 194.A-2003)




\bibliographystyle{mnras}
\bibliography{udspsb} 

\appendix
\section{Comparison between double powerlaw and two component burst fits}\label{App:dblplaw}
In Figure A1 (available online) we show some examples of star formation histories obtained from a double powerlaw and two component burst model. For a galaxy which has undergone a strong recent burst (122200) the double powerlaw is unable to include an old population as the majority of the light is in the short lived intermediate age burst. For a galaxy that appears to undergo a break in star formation (191179) this cannot be modelled by the double powerlaw. For a galaxy with a significant old stellar population as well as a burst (125588) the double powerlaw has to smooth out the burst in order to include a significant enough old stellar population.  

\begin{figure*}
    \centering
    \includegraphics[width=\textwidth]{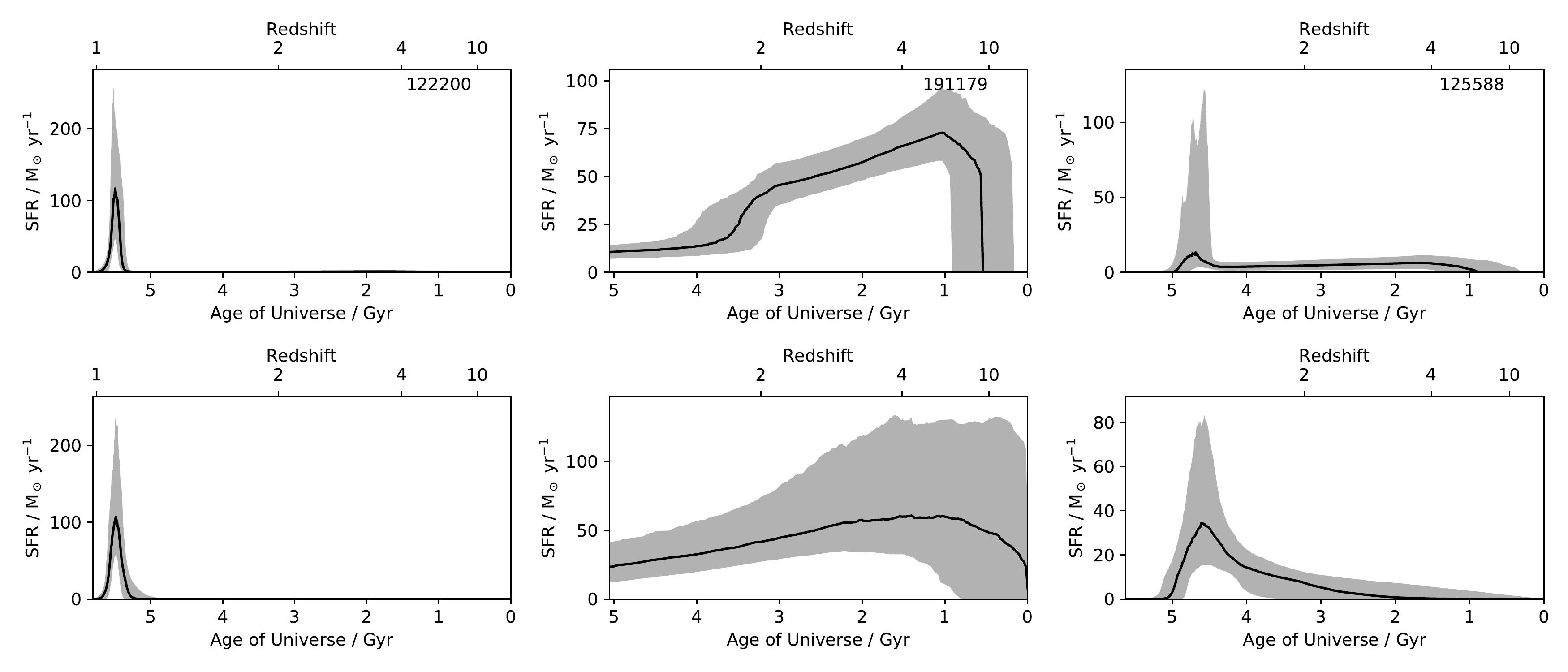}
    \caption{Fitted star formation histories assuming our fiducial exponential+starburst model (top) and initial double powerlaw model (bottom) for 3 example galaxies.}
    \label{fig:dblplaw}
\end{figure*}

\section{Comparison between STARLIGHT and BAGPIPES}\label{App:starlight}

In Figures B1 and B2 (available online) we show comparisons between {\sc starlight} non-parametric fits and {\sc bagpipes} parametric SFH fits to two example objects. There are several important differences between the two codes, beyond just the SFH philosophy. While {\sc bagpipes} is fully Bayesian, {\sc starlight} provides a least squares best-fit model. {\sc starlight} does not include allowance for calibration errors in the spectrophotometry, but does perform a $\sigma$-clipping of extreme outlying pixels in the spectra. We found that this $\sigma$-clipping became extreme when the optical broad-band data was provided to fit alongside the spectra. We therefore exclude these bands from the {\sc starlight} fit, however, they are required in the {\sc bagpipes} fit in order to constrain the spectrophotometric polynomial correction. We additionally did not fit the $3.6\mu$m IRAC band in the {\sc starlight} analysis, as there were concerns at the time about the relative quality of the data compared to the other bands. However, we note that this point is largely superfluous for constraining the shape of the SED at these redshifts where the 2.2$\mu$m K-band is available. 

The top panel of Figures B1 and B2 compare the spectral fits and their residuals. In both cases we plot the best-fit model, for more direct comparison. We see the large-scale bumps and wiggles removed by the Gaussian process noise in the {\sc bagpipes} spectral residuals, however, overall the fits are relatively good for both codes. The middle panels show the photometric data and fitted models. In the case of {\sc bagpipes} we show both the best fit and the posterior distribution. Here we typically see smaller residuals in the {\sc starlight} fit, as expected from the many more degrees of freedom in the star formation and metallicity history. The final row shows the star formation history, for {\sc bagpipes} we show the usual median and percentiles of the posterior distribution, for {\sc starlight} we show the weights given to each simple stellar population (SSP), weighted both by r-band light (top) and mass (bottom). The bars are colour coded by metallicity, while the single metallicity fit by {\sc bagpipes} is written in the left hand panel. In both cases the codes identify a significant recent starburst, although the light-weighted SSPs of {\sc starlight} highlight the difficulty of extracting the amplitude of the old component. While {\sc bagpipes} spreads the old component out over a large time period due to the shape of the assumed prior on the SFH, {\sc starlight} achieves a good fit with just one old component for 157397. For both these galaxies a significant burst is identified, with the {\sc starlight} best-fit fraction of mass formed in the last Gyr within the 1$\sigma$ error bounds of the {\sc bagpipes} fits.

\begin{figure*}
    \centering
    \includegraphics[width=12cm]{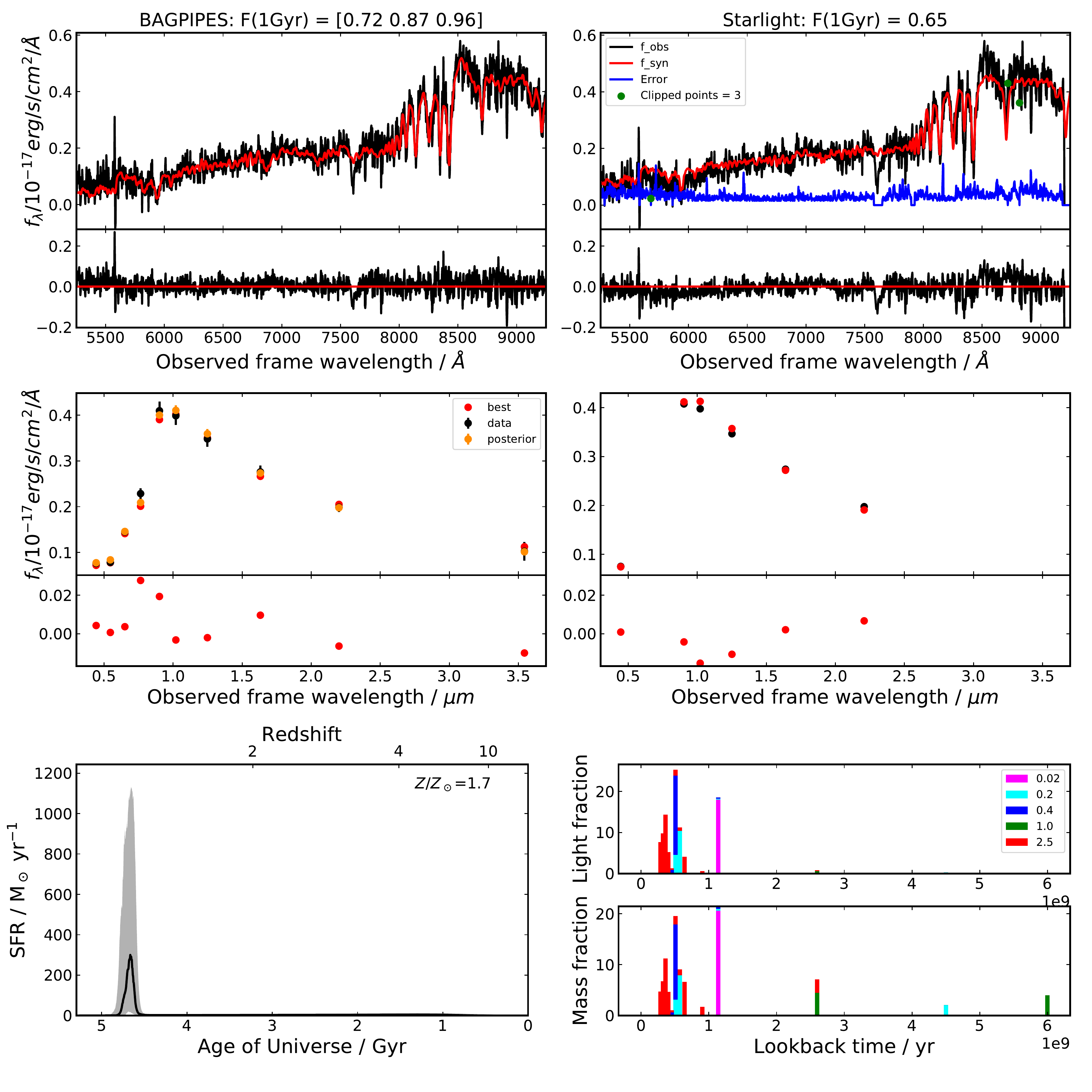}
    \caption{Comparison between spectral fit and resulting star formation history using the {\sc bagpipes} and {\sc starlight} codes for galaxy 108365. The top panels show the observed (black) and best-fit model (red) flux, with the error (blue) in the {\sc starlight} panel, together with the residuals below. The middle panel shows the observed (black) and best-fit model (red) broad band photometry, together with the residuals. For {\sc bagpipes} we also show the posterior fit in orange. The lower panels show the derived star formation history. For {\sc bagpipes} the black line shows the median of the posterior and the grayscale shows the 16th and 84th percentiles. For {\sc starlight} the coloured bars represent the fraction of light (top) and mass (bottom) in each SSP, colour coded by the metallicity of the SSP. The titles provide the fraction of mass formed in the last 1\,Gyr, in the case of {\sc bagpipes} we provide the 16th, 50th and 84th percentiles of the posterior. }
    \label{fig:starlight}
\end{figure*}

\begin{figure*}
    \centering
    \includegraphics[width=12cm]{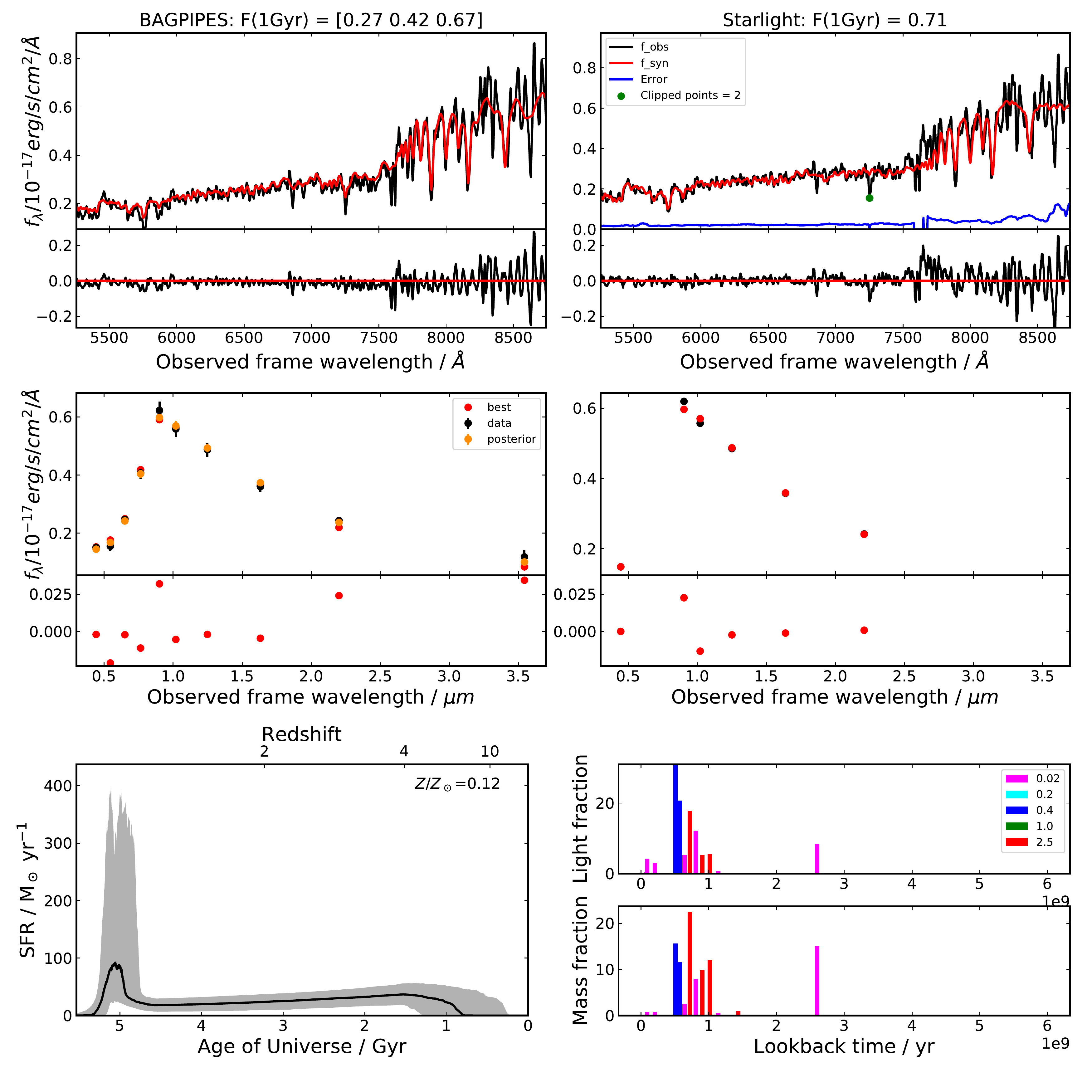}
    \caption{Comparison between spectral fit and resulting star formation history using the {\sc bagpipes} and {\sc starlight} codes for galaxy 157397. See Fig.~ \ref{fig:starlight} for a description. }
    \label{fig:starlight2}
\end{figure*}

\bsp	
\label{lastpage}
\end{document}